\documentclass[11pt]{article}              

\usepackage[margin=20mm]{geometry}
\usepackage{graphicx}
\usepackage{colortbl}
\usepackage[FIGBOTCAP]{subfigure}
 
\usepackage{fancyhdr}
\usepackage{amssymb,amsfonts,amsmath,amsthm}
\usepackage{natbib}

\usepackage{authblk}
	
\graphicspath{{./}}

\usepackage{bbm}
\usepackage{booktabs} 
\usepackage{multirow} 

\usepackage{color}

\usepackage{lineno}

\newcommand{\Rr}{\mathbb{R}}
\newcommand{\cD}{\mathcal{D}}
\newcommand{\ie}{\emph{i.e. }}

\newcommand{\prt}[1]{\left(#1\right)}		
\newcommand{\acc}[1]{\left\{#1\right\}}		
\newcommand{\bra}[1]{\left[ #1 \right]}
			
\newcommand{\Prob}[1]{{\mathbb P}\left( #1 \right)}
\newcommand{\uF}{\overline{F}}
\newcommand{\lF}{\underline{F}}
\newcommand{\ve}[1]{\boldsymbol{#1}}
\newcommand{\vt}{\ve{\theta}}
\newcommand{\vx}{\ve{x}}
\newcommand{\vX}{\ve{X}}
\newcommand{\cm}{\mathcal{M}}
\newcommand{\vC}{\ve{C}}
\newcommand{\vc}{\ve{c}}

\newcommand{\cG}{\mathcal{G}}

\newcommand{\cY}{\mathcal{Y}}
\newcommand{\vCC}{\ve{\mathfrak{C}}}

\newcommand{\vT}{\ve{\Theta}}

\newcommand{\vchi}{\ve{\chi}}

\newcommand{\tr}{^{\textsf T}}
\newcommand{\cs}{\mathcal{S}}
\newcommand{\ub}{\ve{\beta}}
\newcommand{\vtau}{\ve{\tau}}
\newcommand{\vrho}{\ve{\rho}}
\newcommand{\vChi}{\ve{\mathcal{X}}}
\newcommand{\vF}{{\bf F}}
\newcommand{\vR}{{\bf R}}
\newcommand{\muGx}{\mu_{\widehat{Y}}}
\newcommand{\siGx}{\sigma_{\widehat{Y}}}
\newcommand{\lcG}{\underline{\cG}}
\newcommand{\ucG}{\overline{\cG}}
\newcommand{\vTau}{\ve{\mathcal{T}}}

\newcommand{\muMx}{\mu_{\widehat{P}_f}\prt{\vt}}
\newcommand{\siMx}{\sigma_{\widehat{P}_f}\prt{\vt}}

\graphicspath{{./},{./Figures/}}

\begin{document}
\title{Structural reliability analysis for p-boxes using multi-level
    meta-models} 

\author[1]{Roland  Sch\"obi}
\author[1]{Bruno Sudret} 

\affil[1]{Chair of Risk, Safety and Uncertainty Quantification,
    ETH Zurich, Stefano-Franscini-Platz 5, 8093 Zurich, Switzerland}
\date{}
\maketitle

\abstract{In modern engineering, computer simulations are a popular tool to analyse, design, and optimize systems. Furthermore, concepts of uncertainty and the related reliability analysis and robust design are of increasing importance. Hence, an efficient quantification of uncertainty is an important aspect of the engineer's workflow. In this context, the characterization of uncertainty in the input variables is crucial. In this paper, input variables are modelled by probability-boxes, which account for both aleatory and epistemic uncertainty. Two types of probability-boxes are distinguished: free and parametric (also called distributional) p-boxes. The use of probability-boxes generally increases the complexity of structural reliability analyses compared to traditional probabilistic input models. In this paper, the complexity is handled by two-level approaches which use Kriging meta-models with adaptive experimental designs at different levels of the structural reliability analysis. For both types of probability-boxes, the extensive use of meta-models allows for an efficient estimation of the failure probability at a limited number of runs of the performance function. The capabilities of the proposed approaches are illustrated through a benchmark analytical function and two realistic engineering problems.
 \\[1em] 

 {\bf Keywords}: Structural reliability analysis -- Kriging --
 probability-boxes -- design enrichment -- uncertainty quantification
}

\maketitle


\section{Introduction}
\section{Introduction}

Nowadays, computer simulations are a popular engineering tool to design systems of ever increasing complexity and to determine their performance. These simulations aim at reproducing the physical process and hence provide a solution to the underlying governing equations. As an example, finite element models have become a standard tool in modern civil and mechanical engineering. Such models exploit the available computer power, meaning that a single run of the model can take up to hours of computation or more. 

At the same time, uncertainties are intrinsic to these physical processes and to simulation models. Typically, the model parameters are not perfectly known, but inferred from data and then modelled probabilistically. However, a common situation in practice is to have a limited budget for data acquisition and to end up with scarce datasets. This introduces epistemic uncertainty (lack of knowledge) alongside aleatory uncertainty (natural variability). Then, a more general framework is required to characterize the uncertainty in model parameters beyond probability theory. 

\emph{Imprecise probabilities} describes a set of methodologies which are more general than the well-established probability theory. They include Dempster-Shafer's theory of evidence \citep{Dempster1967,Shafer1976}, Bayesian hierarchical models \citep{Gelman2006,Gelman2009}, possibility theory \citep{DeCooman1995,Dubois1988}, probability-boxes \citep{Ferson1996,Ferson2004}, info-gap models \citep{BenHaim2006,Kanno2006}, and fuzzy sets \citep{Zadeh1978,Moller2004}. Among the various methods, probability-boxes provide a simple structure to distinguish aleatory and epistemic uncertainty naturally. 

In this context, engineers are concerned with the reliability of their designs, which is assessed by structural reliability analysis. The goal of such an analysis is the estimation of the failure probability, which describes the probability that the systems performs below a minimal acceptable threshold value. A typical value for the failure probability in engineering is in the range of $10^{-3}$ to $10^{-6}$. Altogether, the complexity of structural reliability analysis comprises three main factors: (i) a high-fidelity computer simulation, (ii) uncertainty in the parameters, and (iii) the estimation of rare events (\ie the estimation of small failure probabilities). 

In the presence of probabilistic variables, a number of methods are known to estimate failure probabilities \citep{Lemaire09,Morio2014}, including Monte Carlo simulation, subset simulation \citep{Au2001,Au2003}, importance sampling, first/second order reliability method (FORM/SORM) \citep{Breitung89,Hohenbichler1988}, and line sampling \citep{Koutsourelakis2004,DeAngelis2015}. In the presence of imprecise probabilities, however, fewer publications are available, some of which are mentioned here. \citep{Eldred2009,Hurtado2013} use nested Monte Carlo simulations to cope with imprecise probabilities. \citep{Alvarez2014} combine subset simulation with random sets. \citep{Zhang2015} generalize FORM/SORM for Dempster-Shafer's evidence theory. A combination of line sampling and imprecise probabilities is discussed in \citep{DeAngelis2015}. These publications typically require a large number of model evaluations and hence rely on an inexpensive-to-evaluate performance function to make the  analysis tractable.

A way to reduce the number of model evaluations and hence allowing expensive-to-evaluate limit-state functions are meta-models. Meta-modelling techniques approximate the expensive-to-evaluate limit-state function by a fast-to-evaluate surrogate based on a relatively small number of runs of the original one. The meta-model is then used in the structural reliability analysis. In particular, meta-models with adaptive experimental designs have been proven to be efficient in the estimation of failure probabilities. As an example, \citep{Bichon2008,Echard2011,Dubourg2013} combine Kriging (Gaussian process models) \citep{Santner2003} and structural reliability analysis in the context of probability theory.

However, only a few authors combine meta-modelling techniques with imprecise probabilities for structural reliability analysis. \citep{Balesdent2014,Morio2016} use an adaptive-Kriging importance sampling algorithm for distributions with interval-valued parameters. A combination of Karhunen-Lo\`eve expansions and evidence theory is presented in \citep{Oberguggenberger2014}.  \citep{Zhang2014} use radial basis expansions in the presence of evidence theory. 

Based on these developments, a set of algorithms is proposed in this paper to further increase the efficiency of structural reliability analysis in the presence of epistemic uncertainties. In particular, free and parametric p-boxes are presented separately and compared later in this paper. The application of Kriging meta-models at several stages of the imprecise structural reliability analysis promises a reduction of the overall computational costs at the same level of accuracy. 

The paper is organized as follows. Section~\ref{sec:pbox} defines probability-boxes as a generalization of probability theory. In Section~\ref{sec:SRA}, the basic structural reliability analysis is set up and Monte Carlo simulation is presented as a solution algorithm. Then, the proposed algorithms to solve imprecise structural reliability analyses are discussed in Section~\ref{sec:isra:free} and \ref{sec:isra:para} for free and parametric probability-boxes, respectively. Finally, the performance of the proposed algorithms is assessed on a benchmark analytical function and realistic engineering problem settings, presented in Section~\ref{sec:appl}.  

\section{Probability-boxes} \label{sec:pbox}
\subsection{Definition}

A probability space is defined as the triplet $(\Omega, \mathcal{F}, \mathbb{P})$, where $\Omega$ is the universal event space, $\mathcal{F}$ is the associated $\sigma$-algebra, and $\mathbb{P}$ is the probability measure. A random variable $X$ defines the mapping $X(\omega):\, \omega\in\Omega\mapsto \mathcal{D}_X\subset\Rr$, where $\omega\in\Omega$ is an elementary event, $\mathcal{D}_X$ is the support of $X$, and $\Rr$ is the set of real numbers. In the context of probability theory, $X$ is typically characterized by its cumulative distribution function (CDF) $F_X$ and, in the case of continuous variables, by its probability density function (PDF) $f_X$. The CDF describes the probability of the event $\acc{X\leq x}$, \ie $F_X(x) = \Prob{X\leq x}$. Its PDF is defined as $f_X(x) = \text{d} F_X(x)/\text{d} x$ and describes the likelihood of $X$ being in the neighbourhood of a given $x\in\cD_X$. 

Probability theory is based on the assumption that the probability measure $\mathbb{P}$ is known precisely, resulting in a well-defined CDF. In cases of incomplete and/or sparse knowledge on $X$, however, aleatory uncertainty (natural variability) and epistemic uncertainty (lack of knowledge) can occur simultaneously. Epistemic uncertainty may not be considered adequately in the context of probability theory alongside aleatory uncertainty \cite{BeerAPSSRA2016}. Hence, a more general formulation of the probability measure, such as in Dempster-Shafer's theory of evidence and probability-boxes, is required to capture the uncertainty fully. 

Dempster-Shafer's theory of evidence accounts for the epistemic uncertainty by replacing the probability measure $\mathbb{P}$ with two values: belief and plausibility \citep{Dempster1967,Shafer1976}. \emph{Belief} measures the minimum amount of probability that must be associated to an event, whereas \emph{plausibility} measures the maximum amount of probability that could be associated to the same event. A \emph{probability-box} (p-box) is a special case of the theory of evidence, considering only the nested set of events $\acc{X \leq x}$, which in turns is related to the definition of a CDF of a probabilistic random variable. 

P-boxes define the CDF of a variable $X$ by lower and upper bounds denoted by $\lF_X$ and $\uF_X$, respectively \citep{Ferson1996,Ferson2004}. For any value $x\in \cD_X$, the true-but-unknown CDF value lies within these bounds such that $\lF_X(x)\leq F_X(x)\leq \uF_X(x)$. The two boundary curves form an intermediate area, hence the name probability-\emph{box}. In the literature, two types of p-boxes are distinguished, namely free and parametric p-boxes, which are discussed in the following.

\subsection{Free p-box} \label{sec:def:free}

\begin{figure}[ht!]
\centering
\subfigure[Free p-box \label{fig:pbox:free}]{
\includegraphics[width = 0.45\linewidth]{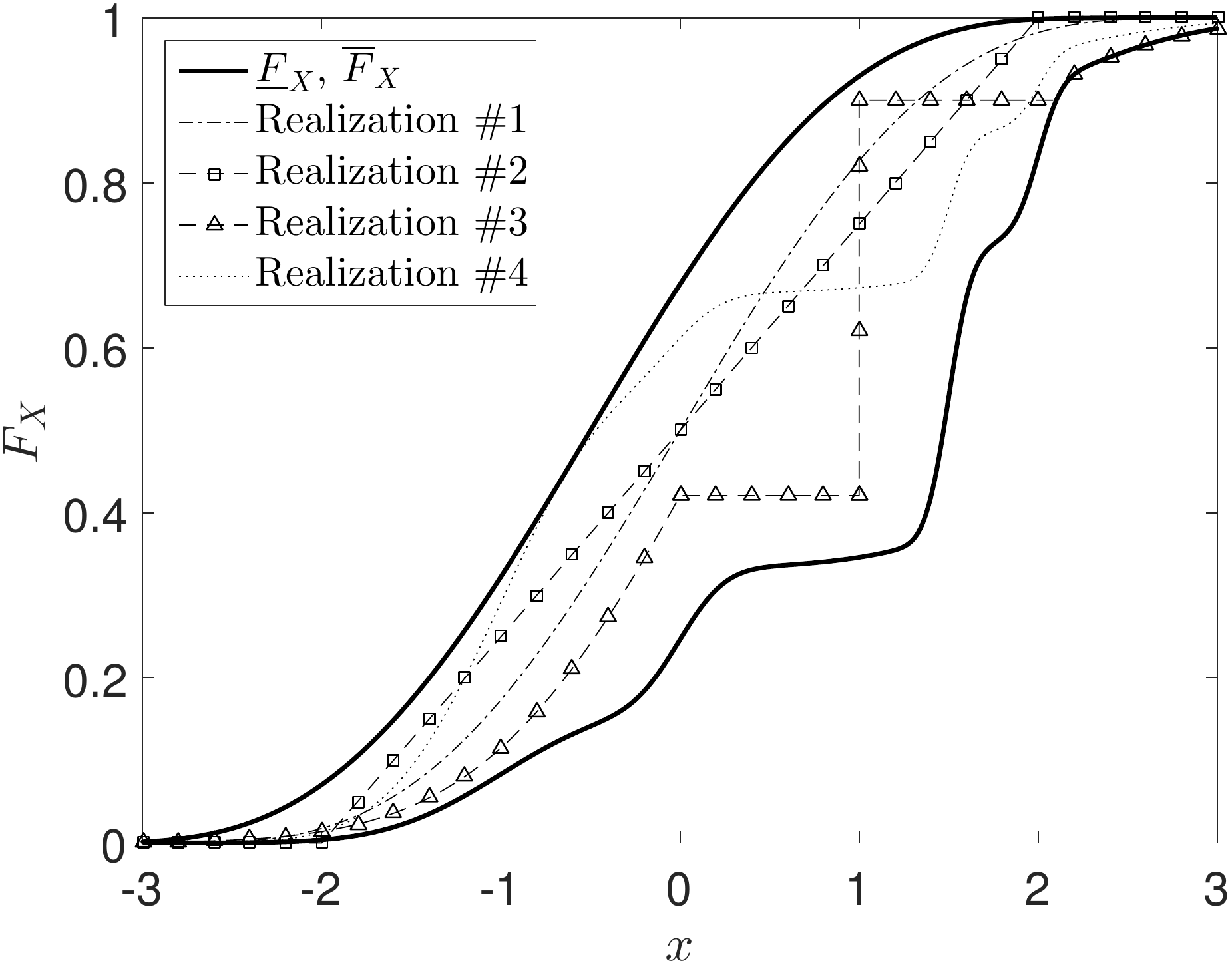}
}
\subfigure[Parametric p-box \label{fig:pbox:para}]{
\includegraphics[width = 0.45\linewidth]{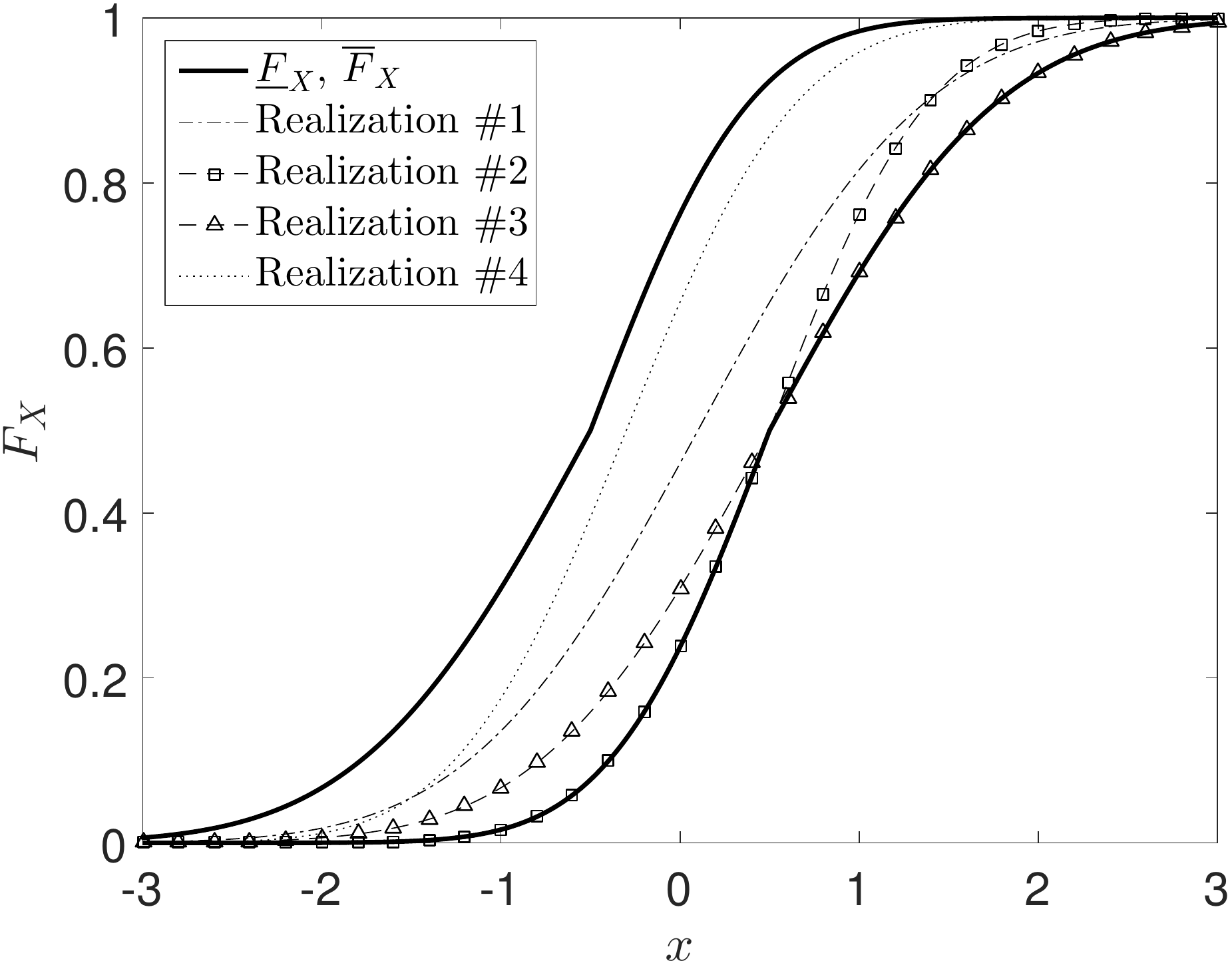}
}
\caption{Definition of p-box types and realizations of possible CDFs \label{fig:pbox}}
\end{figure}

\emph{Free p-boxes} are defined as introduced in the previous section, \ie only by lower and upper bounds of the CDF. This implies that the true CDF can have any arbitrary shape as long as it fulfils the characteristics of a generic CDF and lies within the bounds of the p-box. Figure~\ref{fig:pbox:free} shows the boundary curves of a free p-box and a number of possible realizations of the true CDF. Because the shape of the true CDF is not specified, different types of curves are possible, including non-smooth ones (see realization \#3 in Figure~\ref{fig:pbox:free}).

\subsection{Parametric p-box} \label{sec:def:para}

\emph{Parametric p-boxes} (a.k.a. distributional p-boxes) are defined as cumulative distribution function families the parameters of which are known in intervals:
\begin{equation}
F_X(x) = F_X(x|\vt), \quad \text{s.t.} \quad \vt\in \cD_{\vT}\subset \Rr^{n_{\vt}},
\end{equation}
where $\cD_{\vT}$ denotes the interval domain of the distribution parameters of dimension $n_{\vt}=|\vt|$. When the intervals are independent to each other, $\cD_{\vT}=\bra{\underline{\theta}_1,\overline{\theta}_1} \times \ldots \times \bra{\underline{\theta}_{n_{\vt}},\overline{\theta}_{n_{\vt}}}$ denotes a hyper-rectangular domain.

Parametric p-boxes allow for a clear separation of aleatory and epistemic uncertainty: aleatory uncertainty is represented by the distribution function family, whereas epistemic uncertainty is represented by the intervals in the distribution parameters. However, parametric p-boxes are more restrictive than free p-boxes because of the required knowledge on the distribution family. In other words, free p-boxes can be interpreted as a generalization of parametric p-boxes where the distribution family has $n_{\vt}\rightarrow\infty$ parameters. 

Parametric p-boxes are related to different concepts in the field of imprecise probabilities including Bayesian hierarchical models and fuzzy distributions.  The parametric p-box construction resembles a Bayesian hierarchical model \citep{Gelman2006} in which the distribution of the hyper-parameters $\vt$ is replaced by an interval. From a different point of view, the parametric p-box is a fuzzy distribution function where the membership function of the distribution parameters is equal to one for $\vt\in \cD_{\vT}$ and zero otherwise \citep{Moller2004}. 

Figure~\ref{fig:pbox:para} illustrates a parametric p-box, consisting of a Gaussian distribution family with mean value $\mu\in\bra{-0.5,0.5}$ and standard deviation $\sigma\in\bra{0.7, 1.0}$. The lower and upper boundary curves of the parametric p-box drawn in Figure~\ref{fig:pbox:para} are obtained by: 
\begin{equation}
\lF_X(x) = \min_{\vt\in\cD_{\vT}} F_X(x|\vt), \qquad \uF_X(x) =
 \max_{\vt\in\cD_{\vT}} F_X(x|\vt),
\end{equation}
where $\vt=(\mu,\sigma)\tr$. Hence, the boundary curves shown in Figure~\ref{fig:pbox:para} have the characteristics of a CDF but are not necessarily a realization with a specific parameter combination $\vt_0$. They can consist of sections of different realizations. As an example, the lower boundary CDF is a combination of realization~\#2 ($\mu=0.5$, $\sigma = 0.7$) and realization \#3 ($\mu=0.5$, $\sigma = 1.0$). 

\section{Structural reliability analysis} \label{sec:SRA}
\subsection{Limit-state function}

A limit-state function describes the performance of a process or system as a function of a set of input parameters. The deterministic mapping is defined as:
\begin{equation} \label{eq:lsf}
\cG:\, \vx\in\cD_{\vX}\subset\Rr^M \mapsto y=\cG(\vx)\in\Rr,
\end{equation}
where $\vx$ is a $M$-dimensional vector defined in the input domain $\cD_{\vX}$ and $y$ is the output scalar indicating the performance. The sign of $y$, and hence $\cG(\vx)$, determines whether an input vector $\vx$ corresponds to a safe system ($\cG(\vx)>0$) or a failed system ($\cG(\vx)\leq 0$). The limit-state function $\cG$ is interpreted as a \emph{black box} of which only the input vector $\vx$ and the corresponding response $y=\cG(\vx)$ are accessible. 

\subsection{Failure probability}

Due to uncertainties in the definition of the input vector, $\vX$ is modelled traditionally by probability distributions and in this paper $\vX$ is modelled by p-boxes. Considering a probabilistic input vector $\vX$, the \emph{failure probability} is defined as the probability that the limit-state function takes negative values:
\begin{equation}
P_f = \Prob{Y\leq 0}=\Prob{\cG(\vX)\leq 0},
\end{equation}
which can be recast as an integral:
\begin{equation} \label{eq:int}
P_f = \int_{\cD_f} f_{\vX}(\vx)\text{d}\vx,
\end{equation}
where $\cD_f = \acc{\vx\in\cD_{\vX}:\, \cG(\vx)\leq 0}$ is the failure domain and $f_{\vX}$ is the joint probability density function of the input vector $\vX$. The integration in Eq.~(\ref{eq:int}) cannot be performed analytically in the general case where the failure domain $\cD_f$ has a complex shape. Hence, numerical estimates for the failure probability were developed such as Monte Carlo simulation. Considering a large sample of $\vX$ denoted by $\cs = \acc{\vx_1,\ldots,\vx_n}$, the failure probability can be estimated by:
\begin{equation}
\widehat{P}_f = \frac{n_f}{n} = \frac{1}{n} \sum_{i=1}^{n} \mathbb{I}_{\cG(\vx)\leq 0}(\vx_i),
\end{equation} 
where $n_f$ is the number of failure samples $\vx_i\in\cD_f$, $n=|\cs|$ is the total number of samples and $\mathbb{I}$ is the indicator function with $\mathbb{I} = 1$ for a true subscript statement and $\mathbb{I} =0$ otherwise. Note that the use of $\mathbb{I}$ transforms the structural reliability problem into a classification problem where only the sign of $\cG(\vx)$ is relevant.

In the context of p-boxes, the joint probability density function $f_{\vX}$ is not defined deterministically. Hence, Eq.~(\ref{eq:int}) leads to a range of failure probabilities $P_f\in\bra{\underline{P}_f,\overline{P}_f}$, the bounds of which are defined as:
\begin{equation}\label{eq:int2}
\underline{P}_f = {''}{\min_{f_{\vX}}}'' \int_{\cD_{f}} f_{\vX}(\vx) \text{d}\vx, \qquad \overline{P}_f ={''}{\max_{f_{\vX}}}'' \int_{\cD_{f}} f_{\vX}(\vx) \text{d}\vx,
\end{equation}
where $''{\min}''$ (resp. $''{\max}''$) means that the optimization would be carried out over all PDF $f_{\vX}$ that satisfy some constraints related to the definition of the underlying p-box. 
This \emph{imprecise structural reliability analysis} (ISRA) is not straightforward as it involves an optimization with a multi-dimensional distribution $f_{\vX}$ as the argument of the objective function. In the following sections, solution algorithms for the two types of p-boxes, \ie free and parametric p-boxes, are discussed.

\section{Imprecise structural reliability analysis (ISRA) for free p-boxes} \label{sec:isra:free}
\subsection{Problem conversion}

For the case of free p-boxes, the imprecise structural reliability problem can be recast as two structural reliability problems in the probabilistic sense \citep{Zhang2010,Zhang2012a,SchoebiESREL2015}. These can be solved by conventional structural reliability methods, such as Monte Carlo simulations. 

Consider a random vector $\vC$ of length $M$ which follows a uniform distribution in the unit-hypercube domain $\bra{0,1}^M$. The component $C_i$ shall describe the CDF value of $X_i$. For the sake of simplicity, the input variables $X_i$ are assumed to be statistically independent throughout this paper. When $X_i$ is modelled by a free p-box, each $c_i\in \cD_{C_i}$ can be transformed into an interval in the domain $\cD_{X_i}$ by applying the inverse CDF of the p-box bounds:
\begin{equation}
\underline{x}_i = \overline{F}_{X_i}^{-1}(c_i), \qquad \overline{x}_i = \underline{F}_{X_i}^{-1}(c_i).
\end{equation}
For a given realization of $\vc\in\cD_{\vC}$, let us denote by $\cD_{\vc}=\bra{\underline{x}_1(c_1),\overline{x}_1(c_1)}\times\ldots\times\bra{\underline{x}_M(c_M), \overline{x}_M(c_M)}$. The boundary curves of the p-box of the response variable $Y$ are then obtained by optimization:
\begin{equation} \label{eq:yminymax}
\overline{Y} = \underline{\cG}(\vC) = \min_{\vx\in\cD_{\vc}} \cG(\vx), \qquad
\underline{Y} = \overline{\cG}(\vC) = \max_{\vx\in\cD_{\vc}} \cG(\vx).
\end{equation}
Note that due to the definition of $\ucG$, taking the maximum response value naturally leads to a realization of the lower bound CDF of the response $\underline{Y}$ and vice versa. Based on those boundary curves and the definition of the failure domain $\cD_f$, the range of failure probabilities is obtained by:
\begin{equation} \label{eq:pfminpfmax}
\underline{P}_f = \Prob{\underline{Y}\leq 0} = \Prob{\overline{\cG}(\vC)\leq 0}, \qquad \overline{P}_f = \Prob{\overline{Y}\leq 0} = \Prob{\underline{\cG}(\vC)\leq 0}.
\end{equation}
Then, the bounds of the failure probability can be estimated by sampling the probabilistic input vector $\vC$. Note that the two boundary values can be computed independently.

A drawback of this problem conversion and the subsequent Monte Carlo simulation, however, is the high computational costs due to three distinct causes. Firstly, the computational model can be an expensive-to-evaluate model in many applications, such as finite element models. Secondly, the number of model evaluations may be large due to the optimization operations in Eq.~(\ref{eq:yminymax}), for each $\vc\in\cD_{\vC}$. And thirdly, failure probabilities are generally low, typically $P_f\approx 10^{-3}\ldots 10^{-6}$ in engineering applications, which requires a large number of samples in Monte Carlo simulation to accurately estimate the failure probability, typically $100/P_f$ for a 10~\% accuracy in the estimate. In order to cope with the three main aspects contributing to the total computational effort, meta-models are introduced at different levels of the ISRA for free p-boxes.

\subsection{Adaptive Kriging Monte Carlo simulation}

\subsubsection{Kriging}
Consider again the limit-state function in Eq.~(\ref{eq:lsf}) and a probabilistic input vector $\vX$. Kriging is a meta-modelling approach that considers the limit-state function to be a realization of a Gaussian process \citep{Santner2003}:
\begin{equation}
Y^{(\text{K})}(\vx) = \ub\tr \ve{f}(\vx) + \sigma^2 Z(\vx,\omega),
\end{equation}
where $\ub\tr \ve{f}(\vx)$ is the mean value of the Gaussian process consisting of regression functions $\ve{f}(\vx) = \acc{f_j(\vx),\,j=1,\ldots,p}$ and a vector of coefficients $\ub\in\Rr^p$, $\sigma^2$ is the variance of the Gaussian process, and $Z(\vx,\omega)$ is a zero-mean, unit-variance stationary Gaussian process, characterized by an autocorrelation function $R(|\vx-\vx'|; \vrho)$ and its hyper-parameter(s) $\vrho$.

Training a Kriging meta-model is based on a set of input realizations $\vChi = \acc{\vchi^{(j)}, \, i=1,\ldots,N}$ and the corresponding values of the limit-state function $\cY = \acc{\cY^{(j)} = \cG\prt{\vchi^{(j)}}, \, j=1,\ldots,N}$. Then, the Kriging parameters are obtained by the generalized least-squares solution \citep{Santner2003}:
\begin{equation}
\ub(\vrho) = \prt{\vF\tr\vR^{-1}\vF}^{-1}\vF\tr\vR^{-1}\cY,
\end{equation}
\begin{equation}
\sigma^2_y(\vrho) = \frac{1}{n} \prt{\cY-\vF\ub}\tr\vR^{-1}\prt{\cY-\vF\ub},
\end{equation}
where $\vR_{ij} = R\prt{\left|\vchi^{(i)}-\vchi^{(j)}\right|; \vrho}$ is the correlation matrix and $\vF_{ij} = f_j\prt{\vchi^{(i)}}$ is the information matrix. When the correlation parameters $\vrho$ are unknown, their values can be estimated by maximum likelihood or cross validation \citep{Bachoc2012}. 

Given an arbitrary point $\vx$ of the input domain $\cD_{\vX}$, the prediction value of the limit-state function is a Gaussian variable with mean value and variance:
\begin{equation}
\muGx(\vx) = \ve{f}\tr(\vx)\ub + \ve{r}\tr(\vx) \vR^{-1}\prt{\cY-\vF\ub},
\end{equation}
\begin{equation}
\siGx^2(\vx) = \sigma^2_y\prt{1-\ve{r}\tr(\vx)\vR \ve{r}(\vx) + \ve{u}\tr(\vx) \prt{\vF\tr\vR^{-1}\vF}^{-1} \ve{u}(\vx)},
\end{equation}
where $r_i(\vx) = R\prt{\left|\vx-\vchi^{(i)}\right|; \vrho}$ and $\ve{u}(\vx) = \vF\tr\vR^{-1}\ve{r}(\vx) - \ve{f}(\vx)$. The Kriging predictor interpolates the experimental design, meaning that $\muGx\prt{\vchi^{(i)}} = \cG\prt{\vchi^{(i)}}$ and $\siGx\prt{\vchi^{(i)}}=0$ for $\vchi^{(i)}\in\vChi$. 

\subsubsection{Adaptive experimental designs} \label{sec:free:ak:adaptive}
Kriging is an interpolation algorithm which approximates the limit-state function most accurately close to the points of the experimental design $\vChi$. However, these points are not necessarily suitable to approximate the boundary between the safety and failure domain, \ie the \emph{limit-state surface} $\cG(\vx) = 0$, and hence to estimate the failure probability. Then, an active learning algorithm can improve the estimation of $P_f$ by enriching the experimental design in a guided way. 

\emph{Adaptive Kriging Monte Carlo simulation} (AK-MCS) combines Kriging with an enrichment scheme and Monte Carlo simulation, as proposed in \cite{Echard2011} and further discussed in \cite{SchobiASCE2015,UQdoc_09_107}. The main steps of AK-MCS are described here:
\begin{enumerate}
\item Generate a small initial experimental design $\vChi$ and compute the corresponding response of the limit-state function: $\cY^{(i)} = \cG\prt{\vchi^{(i)}}$.
\item Train a Kriging meta-model $\cG^{(\text{K})}$ based on $\acc{\vChi,\cY}$. In this paper, ordinary Kriging meta-models are used based on a constant (yet unknown) trend $\beta$. 
\item Generate a large set of candidate samples $\cs=\acc{\vx_1,\ldots,\vx_n}$ from $\vX$ and evaluate the response of the meta-model: $\muGx(\vx)$ and $\siGx^2(\vx)$.
\item Compute the failure probability and its confidence values:
\begin{equation}
\widehat{P}_f = \Prob{\muGx(\vx)\leq 0},
\end{equation}
\begin{equation}
\widehat{P}^+_f = \Prob{\muGx(\vx) - k\siGx(\vx)\leq 0}, \qquad \widehat{P}^-_f = \Prob{\muGx(\vx) + k\siGx(\vx)\leq 0}.
\end{equation}
In practice, $k=2$ is selected which approximately corresponds to a point-wise 95~\% confidence interval on $\widehat{Y}\prt{\vx}$.
\item Check the convergence criterion for the estimate of the failure probability:
\begin{equation} \label{eq:stop}
\frac{\widehat{P}_f^+-\widehat{P}_f^-}{\widehat{P}_f}\leq \epsilon_{P_f}.
\end{equation}
It has been shown that $\epsilon_{P_f} = 5\%$ gives accurate results at reasonable costs \citep{SchobiASCE2015}. If the criterion is fulfilled, terminate the adaptive algorithm and return the last meta-model $\cG^{(\text{K})}$. Otherwise, continue with the next step.
\item Compute the probability of misclassification for every $\vx_i\in\cs$ \cite{Bect2012}:
\begin{equation}
P_m(\vx_i) = \Phi\prt{-\frac{\left|\muGx(\vx_i)\right|}{\siGx(\vx)}},
\end{equation}
where $\Phi\prt{\cdot}$ is the CDF of the standard normal variable. Select the best next sample to be added to the experimental design, which maximizes the probability of misclassification:
\begin{equation}
\vchi^* = \arg\max_{i=1,\ldots,n} P_m\prt{\vx_i}.
\end{equation}
\item Add $\vchi^*$ to the experimental design and compute $\cY^* = \cG\prt{\vchi^*}$. Then, go back to step 2 to update the meta-model with the enriched experimental design. 
\end{enumerate}
After the termination of the adaptive Kriging iterations, the failure probability is estimated based on the last meta-model $\cG^{(\text{K})}$ and the Monte Carlo sample $\cs$. 

\subsection{Meta-modelling the limit-state surface $\cG(\vx) = 0$} \label{sec:first}
A first meta-model is applied to the limit-state function $\cG$ and in particular to model the limit-state surface $\acc{\vx\in\cD_{\vX}:\ \cG(\vx)=0}$. However, in order to conduct an AK-MCS analysis, a probabilistic input vector is required. When the input is modelled by free p-boxes in ISRA, an auxiliary input vector $\widetilde{\vX}$ is created by \emph{condensation} \citep{SchoebiJCP2016}. In this paper, the auxiliary input variables $\widetilde{X}_i$ are characterized by the average value of the boundary curves of the p-box:
\begin{equation} \label{eq:aux}
F_{\widetilde{X}_i}(x_i) = \frac{1}{2}\prt{\lF_{X_i}(x_i) + \uF_{X_i}(x_i)}.
\end{equation}
The auxiliary distribution covers the shape of the p-box. Hence, the resulting meta-model $\cG^{\text{(K)}}$ is accurate in the very neighbourhood of the limit-state surface that contributes to the value of the failure probability in the p-box setting. 

\subsection{Meta-modelling the limit-state surfaces $\lcG(\vx) = 0$ and $\ucG(\vx)=0$}
The second meta-model is applied to the limit-state functions $\lcG$ and $\ucG$ and the estimation of the bounds of the failure probability (see Eqs.~(\ref{eq:yminymax}) and (\ref{eq:pfminpfmax})). By using the approximation of $\cG$, Eq.~(\ref{eq:yminymax}) may be replaced by:
\begin{equation} \label{eq:lGuG}
\underline{\cG}(\vc) \approx \min_{\vx\in\cD_{\vc}} \cG^{(\text{K})}(\vx), \qquad
\overline{\cG}(\vc) \approx \max_{\vx\in\cD_{\vc}} \cG^{(\text{K})}(\vx),
\end{equation}
where $\cG^{(\text{K})}$ is the meta-model resulting from the first-level approximation. Therefore, the bounds of the failure probability $\underline{P}_f$ and $\overline{P}_f$ can be estimated by two independent AK-MCS analyses of $\ucG$ and $\lcG$, respectively, \ie $\underline{P}_f=\Prob{\overline{\cG}\prt{\vC}\leq 0}$ and $\overline{P}_f=\Prob{\underline{\cG}\prt{\vC}\leq 0}$ (see also Eq.~(\ref{eq:pfminpfmax})). Note that here the input vector consists of the probabilistic input vector $\vC$, which makes the definition of an auxiliary distribution as in Section~\ref{sec:first} unnecessary. 

To improve the convergence of the AK-MCS algorithms, however, an auxiliary distribution may be beneficial. In fact, the auxiliary distribution $\widetilde{\vX}$ defined in Eq.~(\ref{eq:aux}) is suitable when p-boxes are unbounded, as shown in \cite{SchoebiESREL2015}. Then, an isoprobabilistic transform $T$ maps $\widetilde{\vX}$ to $\vC$ and vice versa. The failure probability is estimated by $\underline{P}_f = \Prob{\overline{\cG}\prt{T\prt{\widetilde{\vX}}}\leq 0}$ and $\overline{P}_f = \Prob{\underline{\cG}\prt{T\prt{\widetilde{\vX}}}\leq 0}$, respectively. 

\subsection{Two-level meta-modelling approach}
\begin{figure}[ht!]
\centering
\includegraphics[width=0.7\linewidth]{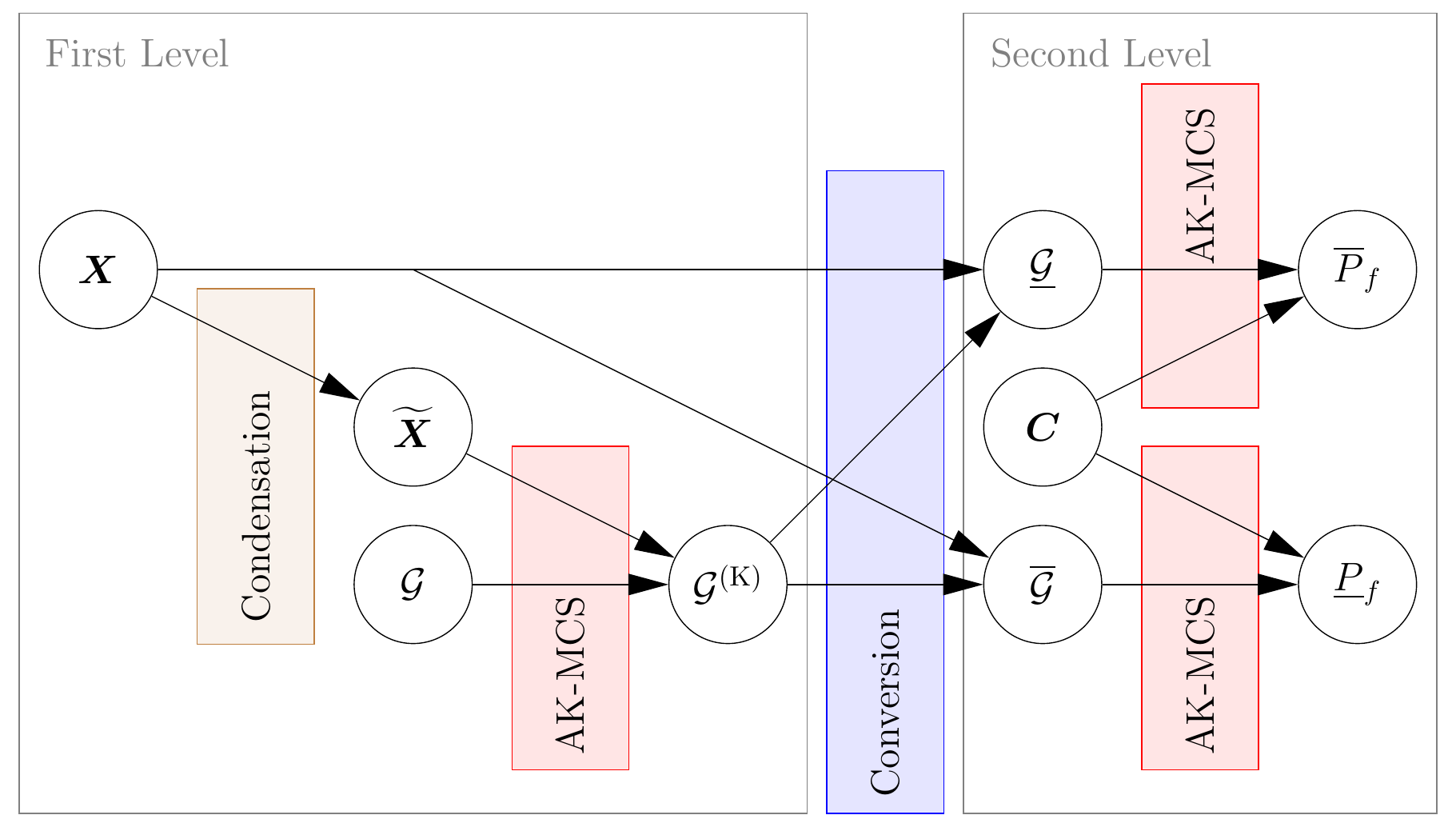}
\caption{Imprecise structural reliability analysis for free p-boxes -- flowchart of the two-level meta-modelling algorithm \label{fig:free:flow}}
\end{figure}

Figure~\ref{fig:free:flow} summarizes the procedure, which consists of two sequential meta-modelling levels connected by a model conversion step. The first-level meta-model surrogates the original limit-state function $\cG$, whereas the two second-level ones surrogate the limit-state functions for estimating the lower and upper bounds of the failure probability. Both levels use auxiliary random variables, \ie $\widetilde{\vX}$ on the first level and $\vC$ on the second level.

\section{Imprecise structural reliability analysis (ISRA) for parametric p-boxes} \label{sec:isra:para}
\subsection{Nested algorithm}

The definition of parametric p-boxes indicates a hierarchical model where the distribution of $\vX$ is defined conditionally on its distribution parameters. Hence, nested simulation algorithms can be applied in the context of ISRA, as discussed in \cite{Eldred2009,SchoebiICASP2015}. In other words, the bounds of the failure probability can be found by estimating the conditional failure probabilities $P_{f|\vt} = \Prob{\cG\prt{\vX_{\vt}}\leq 0}$, where $\vX_{\vt}$ is a conditional distribution with $F_{\vX_{\vt}}(\vx) = F_{\vX}\prt{\vx|\vt}$, and minimizing/maximizing it with respect to $\vt$ to get the bounds $\underline{P}_{f}=\min_{\vt\in\cD_{\vT}} P_{f|\vt}$ and $\overline{P}_{f}=\max_{\vt\in\cD_{\vT}} P_{f|\vt}$.

\begin{figure}[ht!]
\centering
\includegraphics[width = 0.5\linewidth]{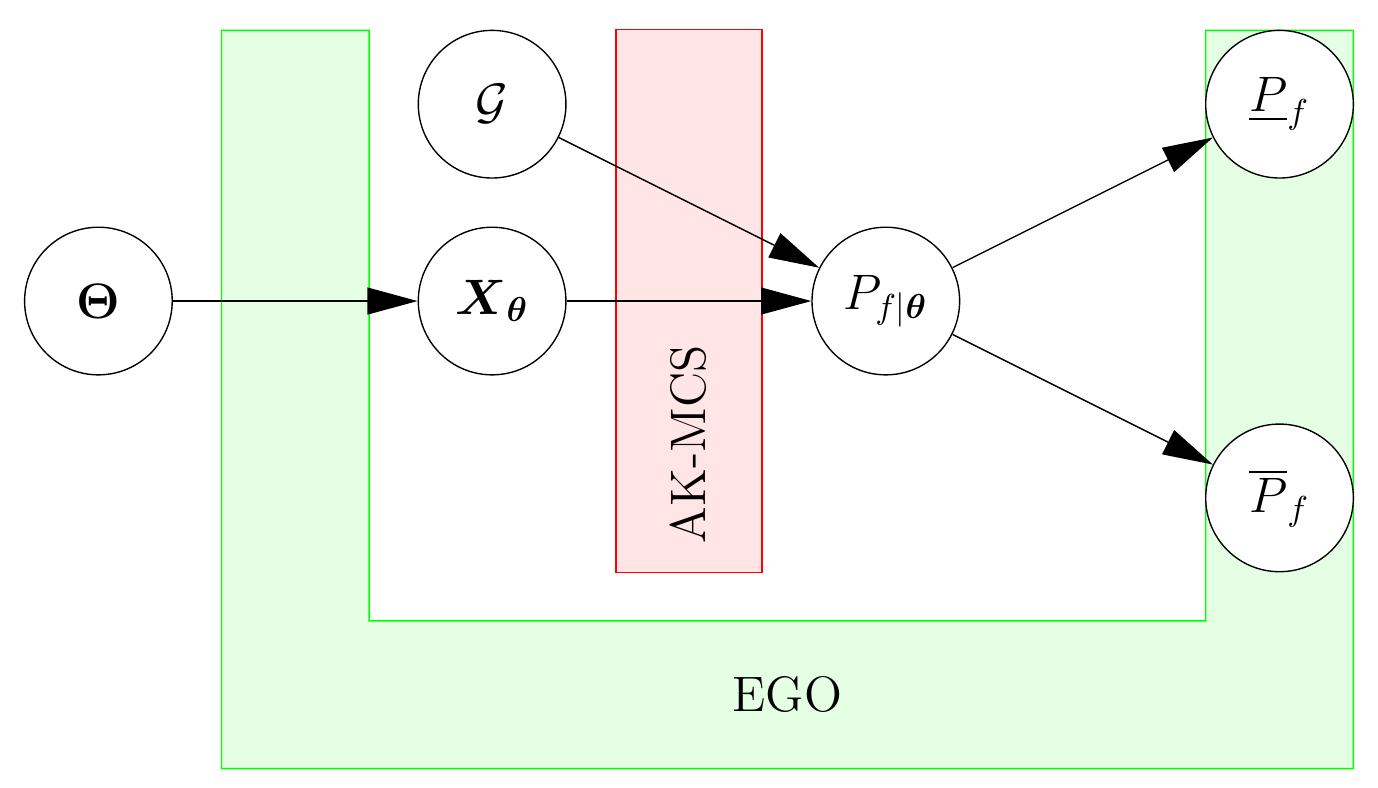}
\caption{Imprecise structural reliability analysis for parametric p-boxes -- overview of the nested algorithm \label{fig:para:flow}}
\end{figure}

Making use of the same tools as in Section~\ref{sec:isra:free} for free p-boxes, the failure probability can be found with the help of Kriging meta-models. The conditional failure probabilities $P_{f|\vt}$ are estimated by AK-MCS, whereas the efficient global optimization (EGO) \cite{Jones1998} algorithm is used to optimize on $\vt$. Figure~\ref{fig:para:flow} illustrates the main components of the proposed algorithm. In the following two sections, AK-MCS and EGO are discussed more in depth.  

\subsection{AK-MCS for conditional failure probabilities}

In the context of parametric p-boxes, the core task consists of estimating efficiently the conditional failure probability $P_{f|\vt}$ by AK-MCS. Given a vector $\vt$ and consequently the conditional variable $\vX_{\vt}$, AK-MCS can be applied to estimate the conditional failure probability as described in Section~\ref{sec:free:ak:adaptive}.

\subsection{Adaptive-Kriging for the lower bound $\underline{P}_f$} \label{sec:ego:min}

EGO is a global optimization algorithm which is based on Kriging meta-models with a design enrichment strategy similar to AK-MCS \citep{Echard2011,Jones1998}. In the context of ISRA for parametric p-boxes, EGO is used to find the extreme values of the failure probability by optimizing on the distribution parameters $\vt$. The main steps for finding the minimum failure probability are discussed here:
\begin{enumerate}
\item Given the distribution parameter domain $\vt\in\cD_{\vT}$, generate a small initial experimental design $\vTau = \acc{\vtau^{(1)},\ldots,\vtau^{(N)}}$. 
\item Compute the corresponding failure probabilities $\mathcal{P}^{(i)}=\Prob{\cG\prt{\vX_{\vtau^{(i)}}}\leq 0}, \, i=1,\ldots,N$ by AK-MCS. Note that for the first evaluation (\ie $\vt=\vtau^{(1)}$) of such a conditional failure probability, the procedure of Section~\ref{sec:free:ak:adaptive} is applicable one-to-one with the distribution $\vX_{\vtau^{(1)}}$. However, for $i=2,\ldots,n$, the limit-state function evaluations of the previous AK-MCS analyses (\ie the samples of the final experimental design $\vChi$) are kept as initial experimental design. The difference between two samples $\vtau^{(i)}$ and $\vtau^{(j)}$ lies then solely in a modified $\vX_{\vtau^{(i)}}$ and hence a modified $\cs_{\vtau^{(i)}}$. Due to the similarity of the MC populations $\cs_{\vtau^{(i)}}$, the number of limit-state function evaluations can be kept small (see also the application in Section~\ref{sec:appl:sdof}). 
\item Train a Kriging meta-model $\cm^{(\text{K})}$ based on $\acc{\vTau,\mathcal{P}}$. This meta-model approximates the estimation of the conditional failure probability $P_{f|\vt}(\vt)$.
\item Search for the optimal samples to be added to the experimental design for minimizing the failure probability:
\begin{equation} \label{eq:vtaumin}
\vtau^{*}_{min} = \arg\max_{\vt\in\cD_{\vT}} \bra{\text{EI}_{min}\prt{\vt}},
\end{equation}
where the expected improvement is defined as \citep{Jones1998,Mockus1978}:
\begin{equation}
\text{EI}_{min}(\vt) = \prt{\mathcal{P}_{min}-\muMx} \ \Phi\prt{\frac{\mathcal{P}_{min}-\muMx}{\siMx}}\ +\ \siMx \ \varphi\prt{\frac{\mathcal{P}_{min}-\muMx}{\siMx}},
\end{equation}
where $\text{EI}_{min}$ is the expected improvement for minimizing  $P_f$, $\mathcal{P}_{min} = \min_i \bra{\mathcal{P}^{(i)}}$ is the current minimum failure probability, $\Phi\prt{\cdot}$ and $\varphi\prt{\cdot}$ are the CDF and PDF of a standard normal variable, respectively. As explained in \cite{Jones1998}, the expected improvement allows for a balance between local and global search.
\item Check the convergence criterion proposed by \cite{Jones1998}, applied to the minimum estimate of the failure probability:
\begin{equation} \label{eq:eimin}
\text{EI}_{min}\prt{\vtau_{min}^*} \leq \epsilon_{EI}.
\end{equation}
If the criterion is fulfilled, terminate the optimization algorithm and return the last estimate of the minimum failure probability: $\underline{P}_f = \min_i\bra{\mathcal{P}^{(i)}}$. Otherwise, continue with Step~6. Note that a value of $\epsilon_{EI}= 10^{-5}$ was identified as a reliable choice for the applications in Section~\ref{sec:appl}.
\item Add the best next sample $\vtau^{*}_{min}$ to the experimental design $\vTau$ according to Eq.~(\ref{eq:vtaumin}).  
\item Evaluate the failure probability corresponding to $\vtau^*_{min}$ and add it to $\mathcal{P}$. 
Note that the limit-state function evaluations computed in previous iterations in AK-MCS can be recycled. They can be used as the initial experimental design in AK-MCS for estimating $\mathcal{P}^*$ corresponding to $\vtau^*$ (see also Step~2 for further details). 
Finally, continue with Step~3. 
\end{enumerate}

\subsection{Adaptive Kriging for the upper bound $\overline{P}_f$} \label{sec:ego:max}
The upper boundary value of the failure probability $\overline{P}_f$ can be estimated by an EGO algorithm analogously to the lower boundary value as follows. The best next sample is determined by replacing Eq.~(\ref{eq:vtaumin}) by:
\begin{equation} \label{eq:vtaumax}
\vtau^{*}_{max} = \arg\max_{\vt\in\cD_{\vT}} \bra{\text{EI}_{max}\prt{\vt}},
\end{equation}
where the expected improvement is defined by:
\begin{equation}
\text{EI}_{max}(\vt) = \prt{\muMx - \mathcal{P}_{max}} \ \Phi\prt{\frac{\muMx - \mathcal{P}_{max}}{\siMx}} \ +\ \siMx \ \varphi\prt{\frac{\muMx - \mathcal{P}_{max}}{\siMx}},
\end{equation}
where $\mathcal{P}_{max} = \max_i \bra{\mathcal{P}^{(i)}}$ is the current maximum failure probability. Furthermore, the algorithm is terminated with a stopping criterion similar to Eq.~(\ref{eq:eimin}):
\begin{equation} \label{eq:eimax}
\text{EI}_{max}\prt{\vtau^*_{max}}\leq \epsilon_{EI}.
\end{equation}
When this is not fulfilled, the best next sample $\vtau^*_{max}$ is added to the experimental design $\vTau$. After termination of the iterative algorithm, the maximum failure probability is estimated by $\overline{P}_f = \max_i\bra{\mathcal{P}^{(i)}}$. 

\subsection{Joint optimization}
The optimization for the minimum and maximum failure probabilities are discussed separately in Sections~\ref{sec:ego:min} and \ref{sec:ego:max}, respectively. However, the Steps~4 and 5 of the EGO algorithm allow for a \emph{simultaneous} optimization for the bounds of the failure probability. For simultaneous optimization, both boundary values of the failure probability are estimated with the same meta-model, $\acc{\vtau^*_{min}, \vtau^*_{max}}$ are determined by Eq.~(\ref{eq:vtaumin}) and (\ref{eq:vtaumax}), respectively, and are added in each iteration of the EGO algorithm. In fact, depending on the two stopping criteria, $\acc{\vtau^*_{min}, \vtau^*_{max}}$ is added if Eqs.~(\ref{eq:eimin}) and (\ref{eq:eimax}) are both \emph{not} fulfilled, $\vtau^*_{min}$ is added if only Eq.~(\ref{eq:eimax}) is fulfilled, or $\vtau^*_{max}$ is added if only Eq.~(\ref{eq:eimin}) is fulfilled. If both equations are fulfilled, the iterative algorithm is terminated. 

For \emph{separate} optimization, the lower and upper bounds of the failure probability are estimated one after the other as presented in the Section~\ref{sec:ego:min} and \ref{sec:ego:max}. The effect of the different optimization modes is discussed in Section~\ref{sec:appl:anal}.


\subsection{Comparison to free p-boxes} \label{sec:para:comp}

ISRA for parametric p-boxes consists of two interconnected Kriging models, the experimental designs of which are enriched quasi simultaneously, as shown in Figure~\ref{fig:para:flow}. In the context of free p-boxes, two levels of meta-models are computed subsequently (see Figure~\ref{fig:free:flow}). Additionally, an optimization operation is required for free p-boxes, which increases the complexity and potentially the computational effort in this case. The increased complexity originates from the basic definition of the free p-box, which can be interpreted as a generalization of the parametric p-box, as discussed in Section~\ref{sec:def:para}. 

Assuming, however, that the computational costs are dominated by the evaluation of the limit-state function, the total computational costs for estimating the boundary values of the failure probabilities may be similar for free and parametric p-boxes. The total number of limit-state function evaluations may be similar due to the same method (\ie AK-MCS) used to surrogate the limit-state surface $\cG(\vx)=0$.

\section{Applications} \label{sec:appl}
\subsection{Two-dimensional toy function} \label{sec:appl:anal}
\subsubsection{Problem statement}

The first example is a simple two-dimensional problem used to illustrate the machinery of the proposed approaches. The limit-state function is defined as:
\begin{equation}
g_1(\vx) = x_1 - x_2^2.
\end{equation}
Failure is defined as $g_1(\vx)\leq 0$ and hence the failure probability is $P_f = \Prob{g_1(\vx)\leq 0}$.  The two variables $X_i$ are modelled by p-boxes. In the following, two cases are distinguished: (i) $X_i$ is modelled by free p-boxes and (ii) $X_i$ is modelled by parametric p-boxes. In the former case, the bounds of the free p-box are defined as:
\begin{equation}
\lF_{X_i}(x_i) = F_{\mathcal{N}}\prt{x_i|\mu=2.5, \sigma=1}, \qquad \uF_{X_i}(x_i) = F_{\mathcal{N}}\prt{x_i|\mu=1.5, \sigma = 1}, \qquad i=1,2,
\end{equation}
where $F_{\mathcal{N}}\prt{x|\mu,\sigma}$ denotes the CDF of a Gaussian distribution with mean value $\mu$ and standard deviation $\sigma$.
For the parametric p-box case, the distribution parameters $\theta$ are chosen so that the boundary curves are the same as in the case of free p-boxes:
\begin{equation}
F_{X_i}(x_i) = F_{\mathcal{N}}\prt{x_i|\mu, \sigma=1}, \quad \mu\in[1.5,\ 2.5].
\end{equation}

\subsubsection{Analysis}

The settings for the ISRA are the following. The meta-models and Monte Carlo simulations are computed using the Matlab-based uncertainty quantification framework {\sc UQLab} \citep{MarelliICVRAM2014}, including its implementation of AK-MCS \citep{UQdoc_09_107}. All Kriging models are defined by a Gaussian autocorrelation function and a constant trend ($f(\vx)=1$, a.k.a. ordinary Kriging). The number of samples in the initial experimental design are $N_1^{0}=12$ and $N_2^{0}=4$ for AK-MCS and EGO, respectively. The initial experimental designs are generated by the Latin-hypercube sampling method \citep{McKay1979} in the unit hypercube $[0,1]^M$ and subsequent isoprobabilistic transform to the corresponding input domains. For free p-boxes, the auxiliary input distribution is defined as $\widetilde{X}_i\sim \mathcal{N}(x_i|2,1)$. For parametric p-boxes and EGO, the threshold value of the stopping criterion is set to $\epsilon_{EI} = 10^{-5}$. The failure probabilities are estimated with a set of $n=10^6$ Monte Carlo samples. The analysis is repeated 50 times with different initial experimental designs in order to obtain statistically significant results. 

\subsubsection{Results}

\begin{figure} 
\centering
\subfigure[Experimental design for $\cG$ (first-level meta-model)\label{fig:anal:free:cg}]{\includegraphics[width=0.45\linewidth]{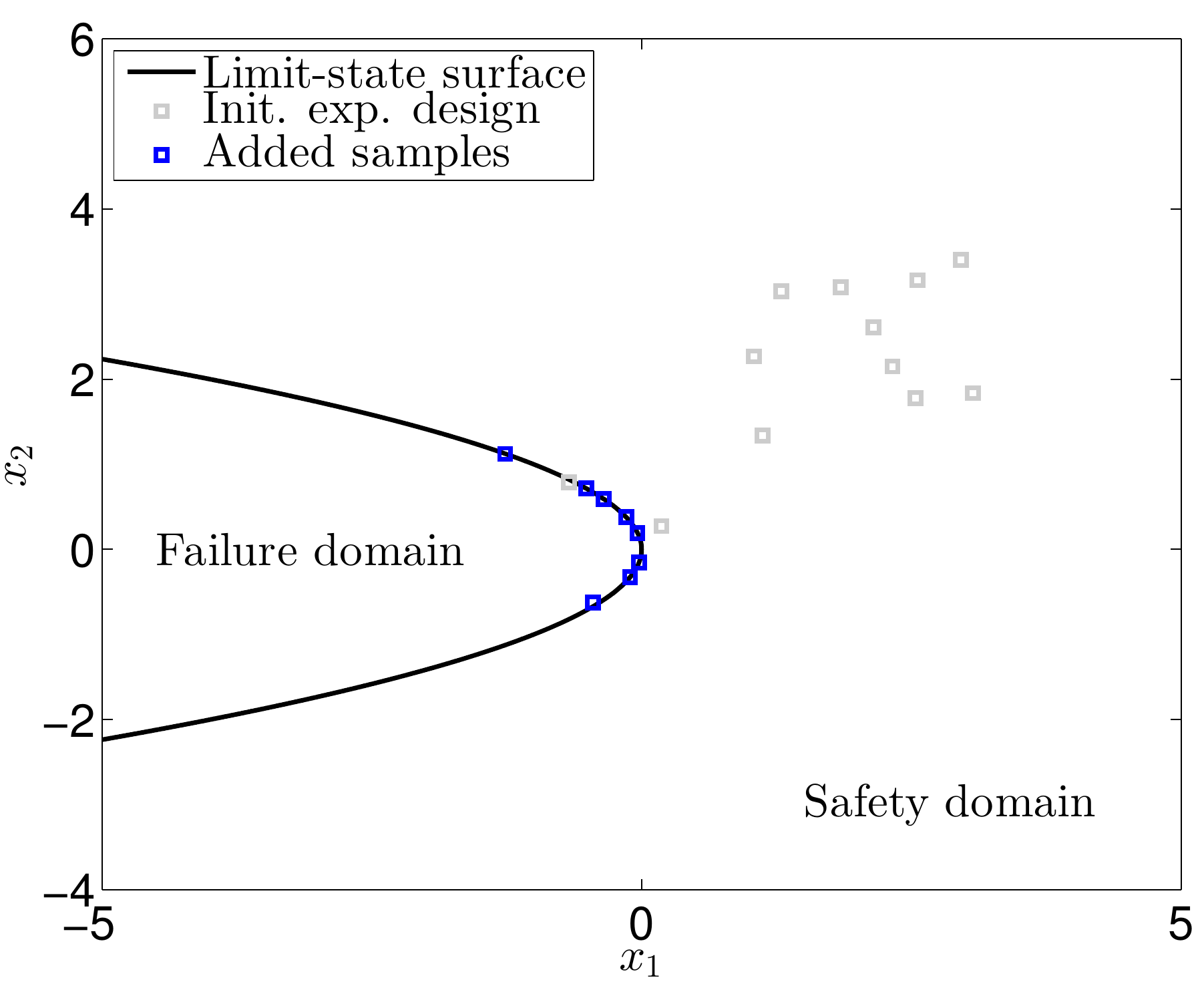}
}\\
\subfigure[Optimization domains in $\cD_{\vX}$ for $\lcG$ (second-level meta-model) \label{fig:anal:free:lcg}]{\includegraphics[width=0.45\linewidth]{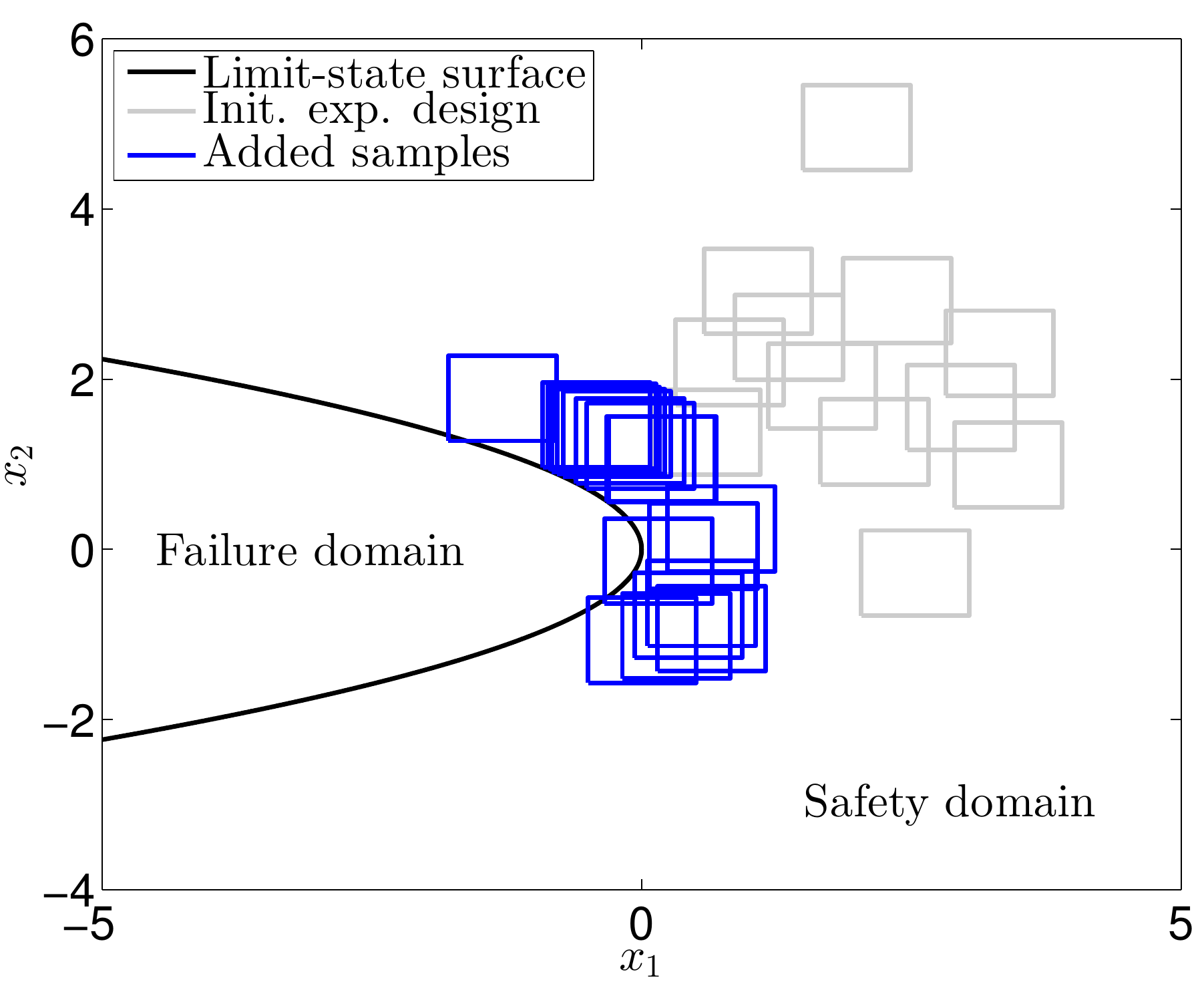}
}
\subfigure[Optimization domains in $\cD_{\vX}$ for $\ucG$ (second-level meta-model) \label{fig:anal:free:ucg}]{\includegraphics[width=0.45\linewidth]{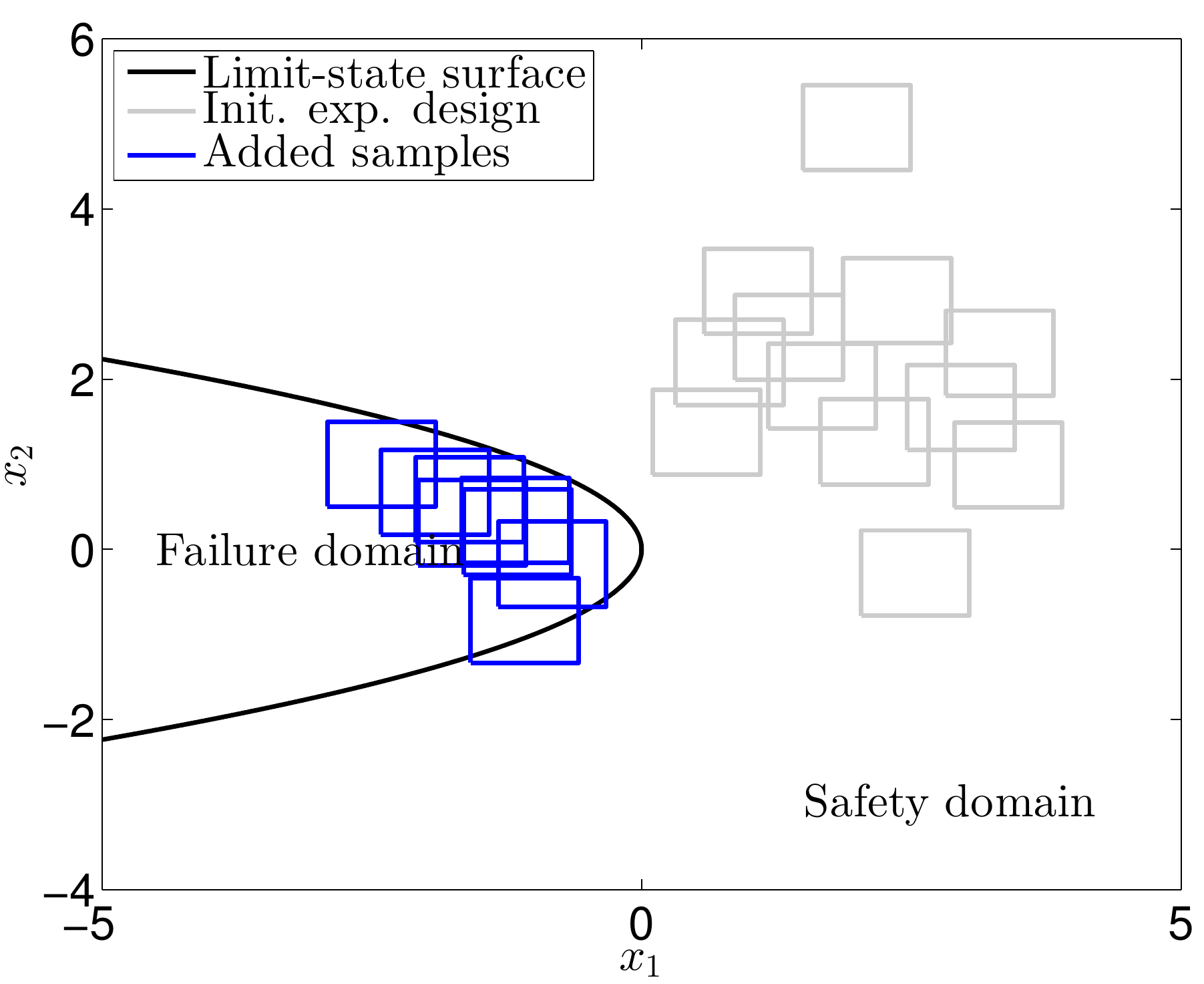}
}
\caption{Two-dimensional toy function -- free p-boxes -- adaptive experimental designs of a single run of the analysis \label{fig:anal:free:all}}
\end{figure}

\paragraph{Free p-boxes}
Figure~\ref{fig:anal:free:all} visualizes a typical realization of the ISRA for free p-boxes.  Note that the stopping criterion in AK-MCS was deactivated in this simulation run to visualize the general behaviour of the method. Figure~\ref{fig:anal:free:cg} illustrates the experimental design of the final first-level meta-model. The initial experimental design is marked by the grey squares. The additional samples (blue squares) group around the limit-state surface (black solid line) in the physical domain, which implies an efficient selection of additional samples. Further, Figures~\ref{fig:anal:free:lcg} and \ref{fig:anal:free:ucg} show the constraint optimization domains $\cD_{\vc}$ in Eq.~(\ref{eq:lGuG}) in the physical space $\mathcal{D}_{\vX}$ for the experimental design $\vCC$ on the second-level meta-model. Again, the optimization domains are selected in an efficient manner, as seen from the positioning of the blue rectangles. For $\underline{\cG}$ in Figure~\ref{fig:anal:free:lcg}, the minimum value of the limit-state function within each blue rectangular is achieved at a point which lies close to the limit-state surface. Thus, the rectangles are mostly contained in the safety domain. Analogously in Figure~\ref{fig:anal:free:ucg} for $\overline{\cG}$, the maximum value of the limit-state function within each blue rectangle is close to the limit-state surface. Hence, the rectangles are mostly contained in the failure domain. 

Accounting for the 50 replications of the ISRA, the estimates of the failure probability are summarized in Table~\ref{tab:anal:free}. The reference value of the failure probability (denoted by $P_{f,ref}$) is obtained by an Importance Sampling analysis of $\underline{\cG}$ and $\overline{\cG}$ with a sample size of $n=10^6$. The table shows that all three meta-models accurately estimate the corresponding failure probabilities in terms of mean estimate. However, in terms of coefficient of variation, the second-level meta-models are less accurate than the expected Monte Carlo sampling-related values: for $\underline{P}_f$ the expected value is $\text{CoV}\bra{\underline{P}_f} = \sqrt{(1-\underline{P}_f)/(n_{MC}\cdot\underline{P}_f)}= 11.3\%< 13.8\%$ and for $\overline{P}_f$ the expected value is $\text{CoV}\bra{\overline{P}_f} = 0.9\%< 7.4\%$. The large variation in $\overline{P}_f$ is caused by the larger number of samples in the failure domain and hence the larger length of the limit-state surface relevant for the failure probability estimation, when comparing to the auxiliary space $\mathcal{D}_{\widetilde{\vX}}$. Selecting a different auxiliary distribution $\widetilde{\vX}$ may possibly reduce this phenomenon. 

\begin{table}[ht!]
\centering
\caption{Two-dimensional toy function -- free p-boxes -- failure probability estimation (results of 50 replications of the same analysis with different sample sets and initial experimental designs) \label{tab:anal:free}}
\begin{tabular}{llll}
\hline
 & $\widetilde{P}_f$ & $\underline{P}_f$ & $\overline{P}_f$ \\
\hline
$P_{f,ref}$ & $1.68\cdot 10^{-3}$ & $7.78\cdot 10^{-5}$ & $1.27\cdot 10^{-2}$\\
$\mathbb{E}\bra{\widehat{P}_f}$ & $1.67\cdot 10^{-3}$ & $7.62\cdot10^{-5}$ & $1.25\cdot 10^{-2}$\\
$\text{CoV}\bra{\widehat{P}_f}$ & $2.5\%$ & $13.8\%$& $7.4\%$\\
\hline
$\mathbb{E}\bra{N_1}$ & $12+2.7=14.7$& - & - \\
$\text{Std}\bra{N_1}$ & $0.6$& - & - \\
$\mathbb{E}\bra{N_2}$ & - & $4+36.9=40.9$& $4+19.4=23.4$ \\
$\text{Std}\bra{N_2}$ & - & $13.3$& $17.7$ \\
\hline
\end{tabular}
\end{table}

The lower part of Table~\ref{tab:anal:free} shows the number of samples in the experimental design of the final meta-models. $N_1$ and $N_2$ denote the number of model evaluations on the first-level and second-level meta-model, respectively (average over 50 replications of the full algorithm).  ${N_1}$ is low due to the smooth shape of the limit-state surface. In fact, an addition of $\Delta N_1=2.7$ points on average is sufficient to model it with sufficient accuracy. On the second level, the number of samples $N_2$ is larger due to the shape of $\underline{\cG}$ and $\overline{\cG}$ and the corresponding limit-state surfaces. 

\begin{figure} 
\centering
\subfigure[Experimental design $\vChi$ for $\cG$ (first-level meta-model)\label{fig:anal:para:doe}]{
\includegraphics[width=0.45\linewidth]{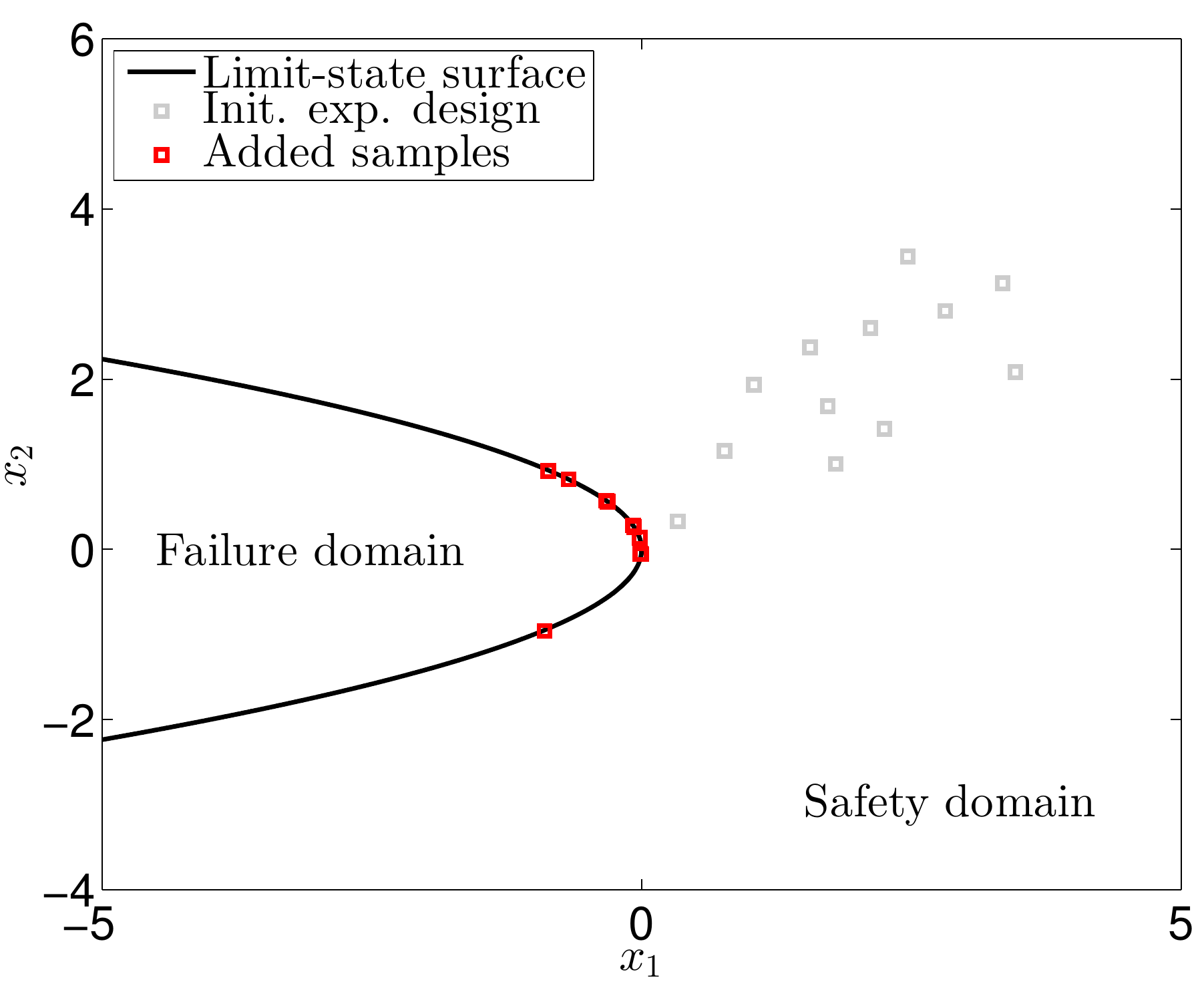}
}\\
\subfigure[Experimental design $\vTau$ in $\mathcal{D}_{\vT}$ (EGO)\label{fig:anal:para:theta}]{
\includegraphics[width=0.45\linewidth]{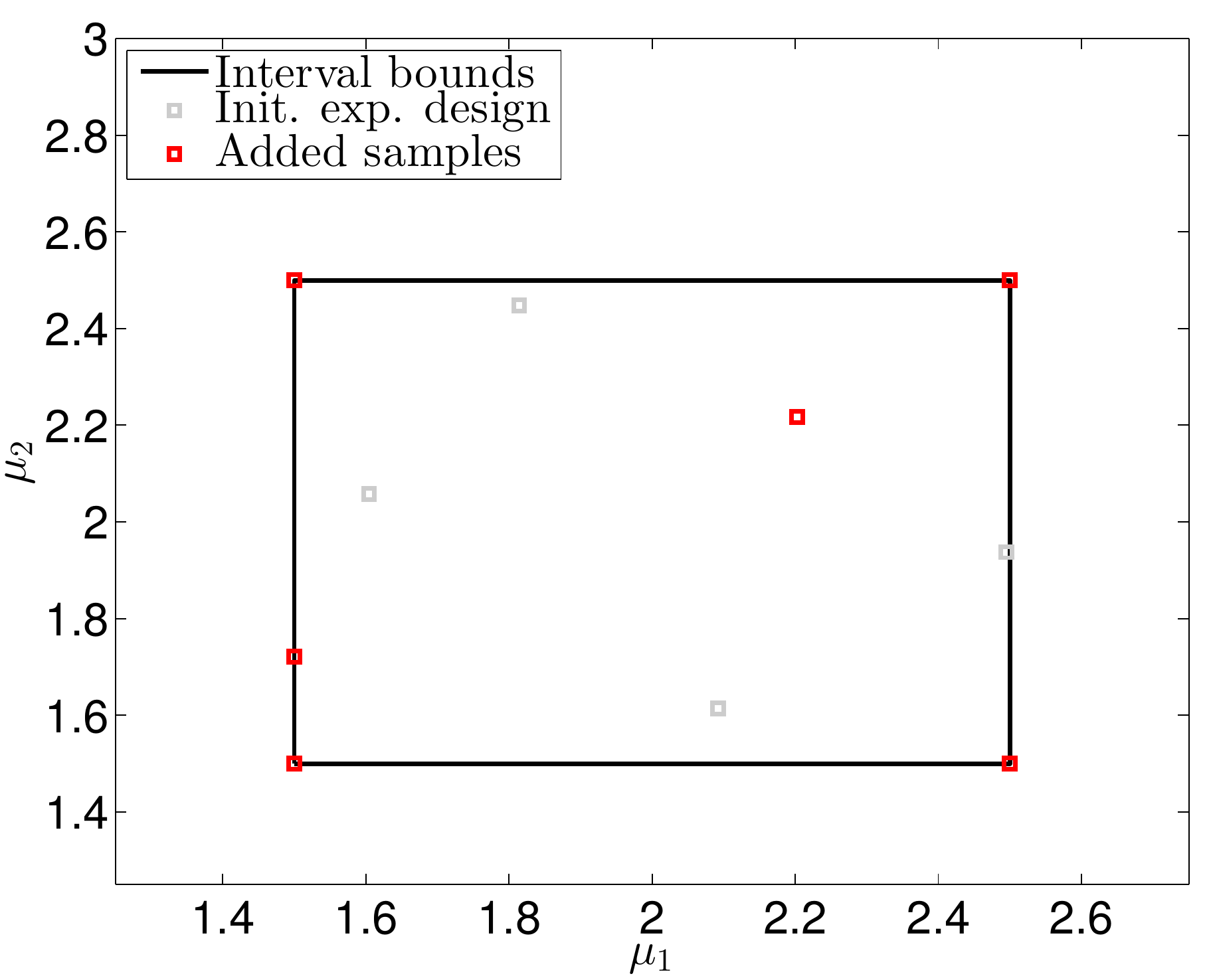}
}
\subfigure[Evolution of the estimate of the failure probability bounds (EGO) \label{fig:anal:para:pf}]{
\includegraphics[width=0.45\linewidth]{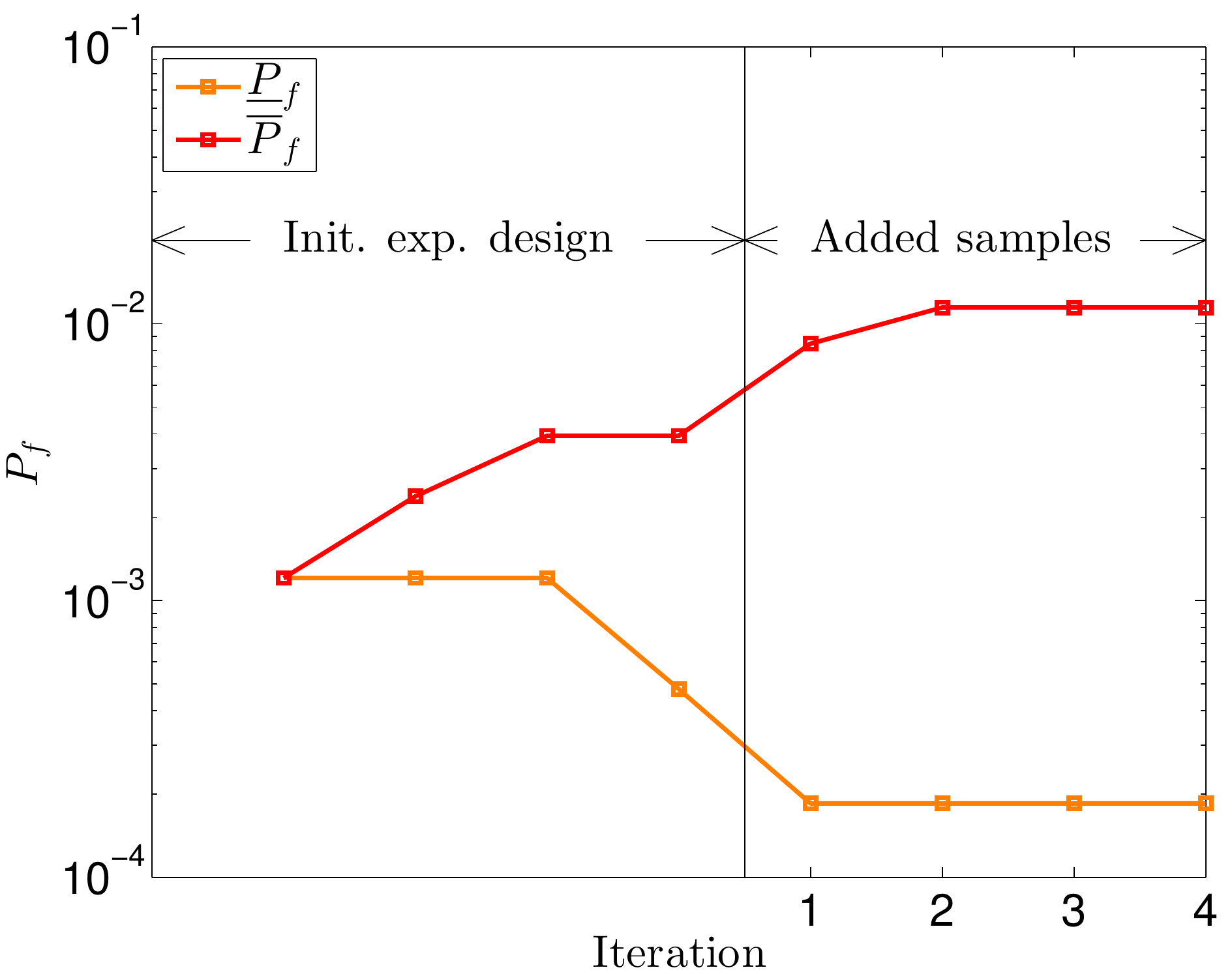}
}
\caption{Two-dimensional toy function -- parametric p-boxes --  single realization of the ISRA \label{fig:anal:para:single}}
\end{figure}

\paragraph{Parametric p-boxes}
A typical realization of the ISRA for parametric p-boxes is visualized in Figure~\ref{fig:anal:para:single}. Figure~\ref{fig:anal:para:doe} shows the experimental design of the final AK-MCS meta-model. Similar to Figure~\ref{fig:anal:free:cg}, the additional samples (red squares) group efficiently around the limit-state surface which is represented by the solid black line. The experimental design of the EGO algorithm is shown in Figure~\ref{fig:anal:para:theta}, which illustrates the exploratory behaviour of the EGO algorithm. Despite the small number of added samples, the boundary values of the failure probability are estimated accurately, as seen in Figure~\ref{fig:anal:para:pf}. The left side of Figure~\ref{fig:anal:para:pf} shows the evolution of the bounds $\bra{\underline{P}_f,\overline{P}_f}$ of the failure probability using the initial experimental design (\ie $N_1^0=4$ samples in $\vChi^0$ are generated simultaneously and the corresponding $\cY^{(i)}$ are computed sequentially), whereas the right side shows them during each iteration of EGO. After iteration~3, the optimization algorithm converges to a stable value of the failure probability bounds. 

The results of the 50 replications of the analysis are summarized in Table~\ref{tab:anal:para} in terms of failure probability estimates and experimental design sizes. Further, it is distinguished whether the EGO algorithm is used to simultaneously optimize for $\underline{P}_f$ and $\overline{P}_f$ (left block of values) or whether EGO is performed separately for  $\underline{P}_f$ and $\overline{P}_f$ (right block of values). Note that for the separate optimization case, samples from one optimization are \emph{not} re-used in the other one. In terms of the failure probability, both settings result in accurate estimates. Moreover, the coefficient of variation is close to the expected value of $\text{CoV}\bra{{P}_f}=8.0\%$ and $0.9\%$ for $\underline{P}_f$ and $\overline{P}_f$, respectively, for a Monte Carlo simulation with $n=10^6$ samples.

\begin{table}[ht!]
\centering
\caption{Two-dimensional toy function -- parametric p-boxes -- failure probability estimation (results of 50 replications of the same analysis with different sample sets and initial experimental designs) \label{tab:anal:para}}
\begin{tabular}{llllll}
\hline
& \multicolumn{2}{c}{Simultaneous optimization} &$\qquad$& \multicolumn{2}{c}{Separate optimization} \\
& $\underline{P}_f$ & $\overline{P}_f$ && $\underline{P}_f$ & $\overline{P}_f$ \\
\hline
$P_{f,ref}$ & $1.57\cdot 10^{-4}$ & $1.14\cdot 10^{-2}$ && $1.57\cdot 10^{-4}$ & $1.14\cdot 10^{-2}$\\
$\mathbb{E}\bra{\widehat{P}_f}$ & $1.58\cdot 10^{-4}$& $1.14\cdot 10^{-2}$ && $1.58\cdot 10^{-4}$ & $1.14\cdot 10^{-2}$\\
$\text{CoV}\bra{\widehat{P}_f}$ & $8.0\%$& $1.2\%$&& $8.1\%$& $1.2\%$\\
\hline
$\mathbb{E}\bra{N_1}$ & \multicolumn{2}{c}{$12+3.4=15.4$} && $12+3.4=15.4$& $12+3.2=15.2$\\
$\text{Std}\bra{N_1}$ & \multicolumn{2}{c}{$0.9$} &&$1.0$ & $0.9$\\
$\mathbb{E}\bra{N_{2}}$ & \multicolumn{2}{c}{$4+10.4=14.3$} && $4+4.7=8.7$& $4+2.0=6.0$\\
$\text{Std}\bra{N_{2}}$ & \multicolumn{2}{c}{$1.4$} && $1.2$& $1.1$\\
\hline
\end{tabular}
\end{table}

Further, the number of evaluations of the true limit-state function $N_1$ is similar for the simultaneous and separate optimization cases. The number of samples added during AK-MCS is generally low. At the level of EGO, however, the simultaneous optimization requires more samples ($N_2$) as the separate optimization. Even when adding up the samples for the separate optimization case, $N_2^{tot} = 4+4.7 + 2.0 = 10.7$ is less than $N_2=14.3$. Hence, the separate optimization is more efficient than the simultaneous optimization in this case. This behaviour has been also observed on other toy functions. Based on this observation, the remaining application examples are analysed using separate optimizations. 

\paragraph{Comparison}
The algorithms for free and parametric p-boxes originate from two different paradigms: interval analysis and nested Monte Carlo simulations, respectively. However, the cost statistics shows that the two proposed two-level meta-modelling algorithms perform similarly in terms of model evaluations. Free p-boxes required $N_1=14.7$ evaluations of the limit-state function whereas parametric p-boxes required $N_1=15.4$ on average. This can be intuitively understood due to the identical limit-state surface in both settings. Contrarily to $N_1$, the number of evaluations on the second-level meta-models, \ie $N_2$, are not comparable for the case of free and parametric p-boxes, due to the different algorithms used.

\subsection{Single-degree-of-freedom oscillator} \label{sec:appl:sdof}

\subsubsection{Problem setup}
\begin{figure}[ht!]
\centering
\includegraphics[width=0.7\linewidth]{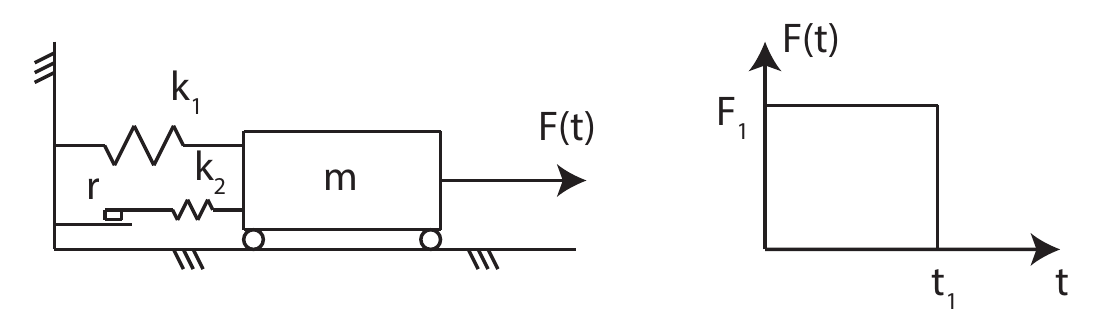}
\caption{\label{fig:osci:sketch} SDOF oscillator -- geometry sketch and definition of the variables}
\end{figure}

In order to visualize the effects of non-monotone functions on the failure probability, the following example is analysed. Consider the non-linear undamped single-degree-of-freedom (SDOF) oscillator sketched in Figure~\ref{fig:osci:sketch} \citep{Echard2011,Echard2013,Schueremans2005}. The corresponding limit-state function reads:
\begin{equation}
g_{SDOF}\prt{r,F_1,t_1,k_1,k_2,m} = 3r - \left| \frac{2 F_1}{m\omega_0^2}\sin\prt{\frac{\omega_0T_1}{2}} \right|,
\end{equation}
where $m$ is the mass, $k_1, k_2$ are the spring constants of the primary and secondary springs, $r$ is the displacement at which the secondary spring yields, $t_1$ is the duration of the loading, $F_1$ is the amplitude of the force and $\omega_0=\sqrt{\frac{k_1+k_2}{m}}$ is the natural frequency of the oscillator. Failure is defined as $g_{SDOF}\leq 0$ and the associated failure probability is $P_f = \Prob{g_{SDOF}\leq 0}$.

The input vector is modelled by a mix of probabilistic variables and p-boxes accounting for the different levels of knowledge. The description of the input variables is provided in Table~\ref{tab:osci:variables}. It is assumed that the spring stiffnesses and the mass are known well. Hence $\acc{k_1,k_2,m}$ are modelled by precise CDFs. On the other side, knowledge on $\acc{r,F_1,t_1}$ is scarce. Hence, these variables are modelled by p-boxes. The two cases of free and parametric p-boxes are distinguished and compared in the following. As seen in Table~\ref{tab:osci:variables}, the parametric p-box is characterized by a distribution function with interval-valued mean value. For the case of free p-boxes, the same  p-box boundary curves are used as for parametric p-boxes. Note that $F_{\mathcal{N}}(x|\mu,\sigma)$ refers to a Gaussian distribution with mean value $\mu$ and standard deviation $\sigma$. 

\begin{table}[ht!]
\centering
\caption{Oscillator -- definition of the input vector for the cases of parametric and free p-boxes \label{tab:osci:variables}}
\begin{tabular}{llllllll}
\hline
& &$\qquad$& \multicolumn{2}{c}{Parametric p-box} &$\qquad$& \multicolumn{2}{c}{Free p-box} \\
Variable & Distribution && Mean & Std. dev. && $\underline{F}_{X_i}$ & $\overline{F}_{X_i}$\\
\hline
$r$ & Gaussian && $[0.49,0.51]$ & $0.05$ && $F_{\mathcal{N}}(r|0.51,0.05)$ & $F_{\mathcal{N}}(r|0.49,0.05)$ \\
$F_1$ & Gaussian && $[-0.2,0.2]$& $0.5$ &&$F_{\mathcal{N}}(F_1|0.2,0.5)$ & $F_{\mathcal{N}}(F_1|-0.2,0.5)$ \\
$t_1$ & Gaussian && $[0.95,1.05]$& $0.2$ && $F_{\mathcal{N}}(t_1|1.05,0.2)$ & $F_{\mathcal{N}}(t_1|0.95,0.2)$\\
\hline
$k_1$ & Gaussian && $1$& $0.1$ && \multicolumn{2}{c}{$F_{\mathcal{N}}(k_1|1,0.1)$}\\ 
$k_2$ & Gaussian && $0.1$& $0.01$ && \multicolumn{2}{c}{$F_{\mathcal{N}}(k_2|0.1,0.01)$} \\ 
$m$ & Gaussian && $1$& $0.05$ && \multicolumn{2}{c}{$F_{\mathcal{N}}(m|1,0.05)$} \\ 
\hline
\end{tabular}
\end{table}

\subsubsection{Analysis}
The settings for the imprecise structural reliability analyses are kept the same as in Section~\ref{sec:appl:anal}, except for the size of the initial experimental designs. The size of the initial experimental designs for all meta-models (including EGO) is set to $N^0=12$. Further, the threshold value in EI is set to $\epsilon_{EI}=10^{-5}$. For free p-boxes, the auxiliary distributions for $\acc{r,F_1,t_1}$ are defined by $F_{\mathcal{N}}(r|0.5,0.05)$, $F_{\mathcal{N}}(F_1|0,0.5)$, and $F_{\mathcal{N}}(t_1|1.00,0.2)$, respectively. 

\subsubsection{Results}
\paragraph{Efficiency of the proposed approaches}
Table~\ref{tab:osci:results} summarizes the results of a single run of the ISRA. The estimates of the failure probability $\widehat{P}_f$ are compared to a reference Monte Carlo simulation obtained with $n_{MC}= 10^7$ samples and the exact limit-state function. The corresponding optimizations are using \textsc{Matlab}'s genetic algorithm. All four failure probabilities are estimated accurately. Further, the number of evaluations of the limit-state function $N_1$ is low. The case of parametric p-boxes results in a larger $N_1=197$ compared to the free p-box ($N_1=132$). The reason for this is the added samples during the iterations of the EGO. In comparison, the two levels of the free p-box analysis are independent and hence the total $N_1$ is lower. 

\begin{table}[ht!]
\centering
\caption{\label{tab:osci:results} Oscillator -- results of the ISRA (reference solution obtained by Monte Carlo simulation with $n_{MC}=10^7$ samples)}
\begin{tabular}{lllllll}
\hline
 &$\quad$& \multicolumn{2}{c}{Parametric p-box} &$\quad$& \multicolumn{2}{c}{Free p-box} \\
 && $\underline{P}_f$ & $\overline{P}_f$ && $\underline{P}_f$ & $\overline{P}_f$ \\
\hline
$P_{f,ref}$ && $2.42\cdot 10^{-3}$ &$9.04\cdot 10^{-3}$&& $7.08\cdot 10^{-4}$ & $1.63\cdot 10^{-2}$ \\
$\widehat{P}_f$ && $2.41\cdot 10^{-3}$ & $9.04\cdot10^{-3}$ && $6.69\cdot 10^{-4}$ & $1.61\cdot 10^{-2}$  \\
$N_1$  && $12+185=197$ & $12+185=197$&& \multicolumn{2}{c}{$12+120=132$}  \\
$N_2$ && $6+3=9$ & $6+6=12$ && $12+60=82$ & $12+153=165$  \\
\hline
$\mu_r^*$ && $1.02$ &$0.98$ &&-&- \\
$\mu_{F_1}^*$ && $-0.01$& $0.20$ &&-&- \\
$\mu_{t_1}^*$ &&$0.95$ & $1.05$ &&-&- \\
\hline
\end{tabular}
\end{table}

The number of samples in the second-level meta-models are varying more than the first-level meta-models. In the case of parametric p-boxes, a few samples $N_2=3\ldots6$ are sufficient to find the optimal distribution parameters. In the case of free p-boxes, $N_2$ has the same order of magnitude as $N_1$ because the analysis is of the same type, \ie AK-MCS. However, the difference in $N_2$ of estimating $\underline{P}_f$ and $\overline{P}_f$ is large, which represents the different complexity of the limit-state surfaces $\ucG=0$ and $\lcG=0$, respectively. 

For the case of parametric p-boxes, Table~\ref{tab:osci:results} shows also the optimal mean values for $\acc{r, F_1, t_1}$, denoted by $\acc{\mu_r^*, \mu_{F_1}^*, \mu_{t_1}^*}$, which result from the EGO optimization. The results confirm the intuition: the largest maximal failure probability is obtained by the smallest yield displacement in combination with a long activation time and an eccentric force ($\mu_{F_1}\approx0.2$ or $\mu_{F_1}\approx -0.2$). On the other hand, the minimum failure probability is obtained by the largest yield displacement in combination with a short activation time and a concentric force ($\mu_{F_1}\approx 0$).

\begin{figure}[ht!]
\centering
\subfigure[Failure probability estimates \label{fig:osci:para:pf}]{
\includegraphics[width=0.45\linewidth]{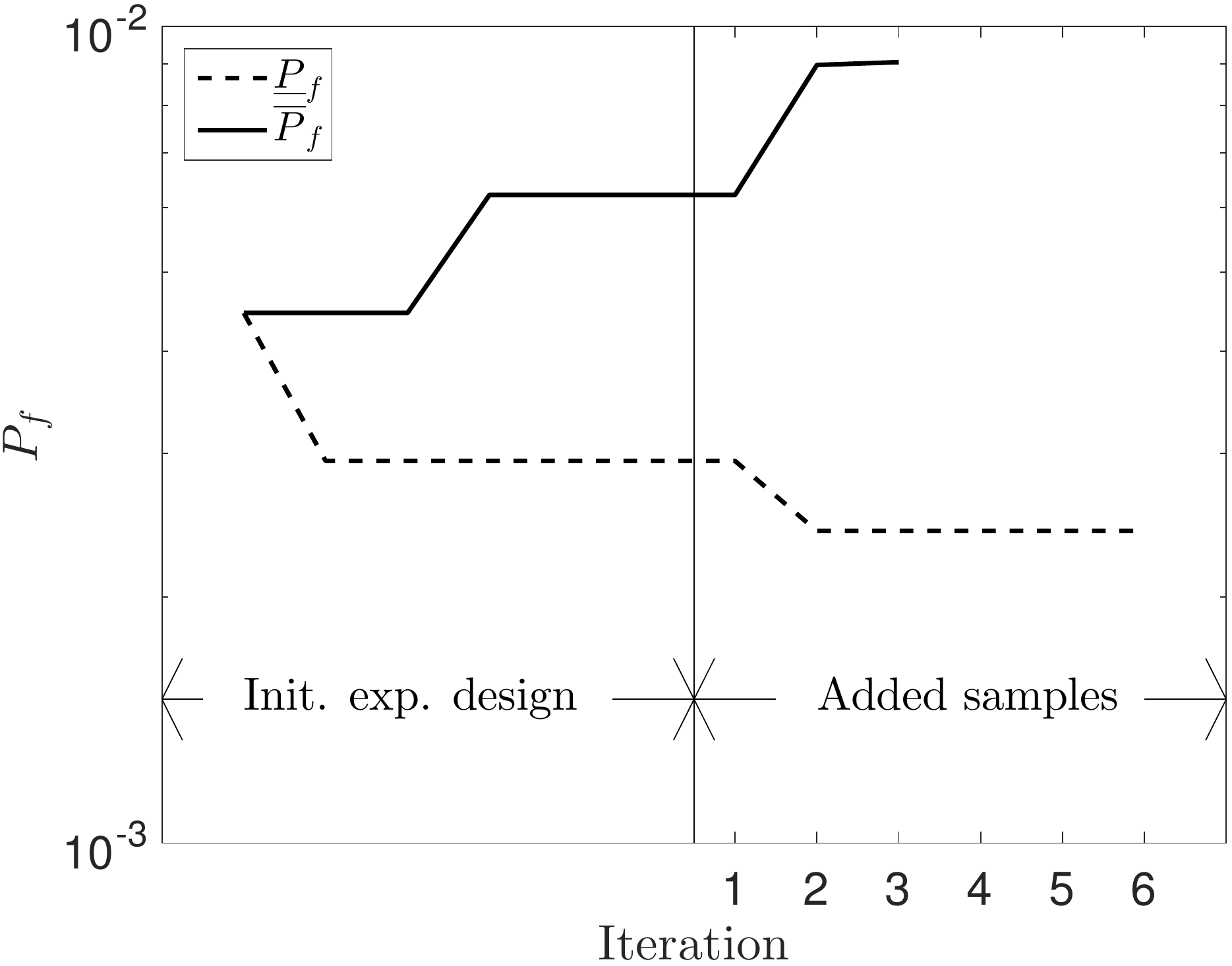}
}
\subfigure[Experimental design size $N_1$ \label{fig:osci:para:doe}]{
\includegraphics[width=0.45\linewidth]{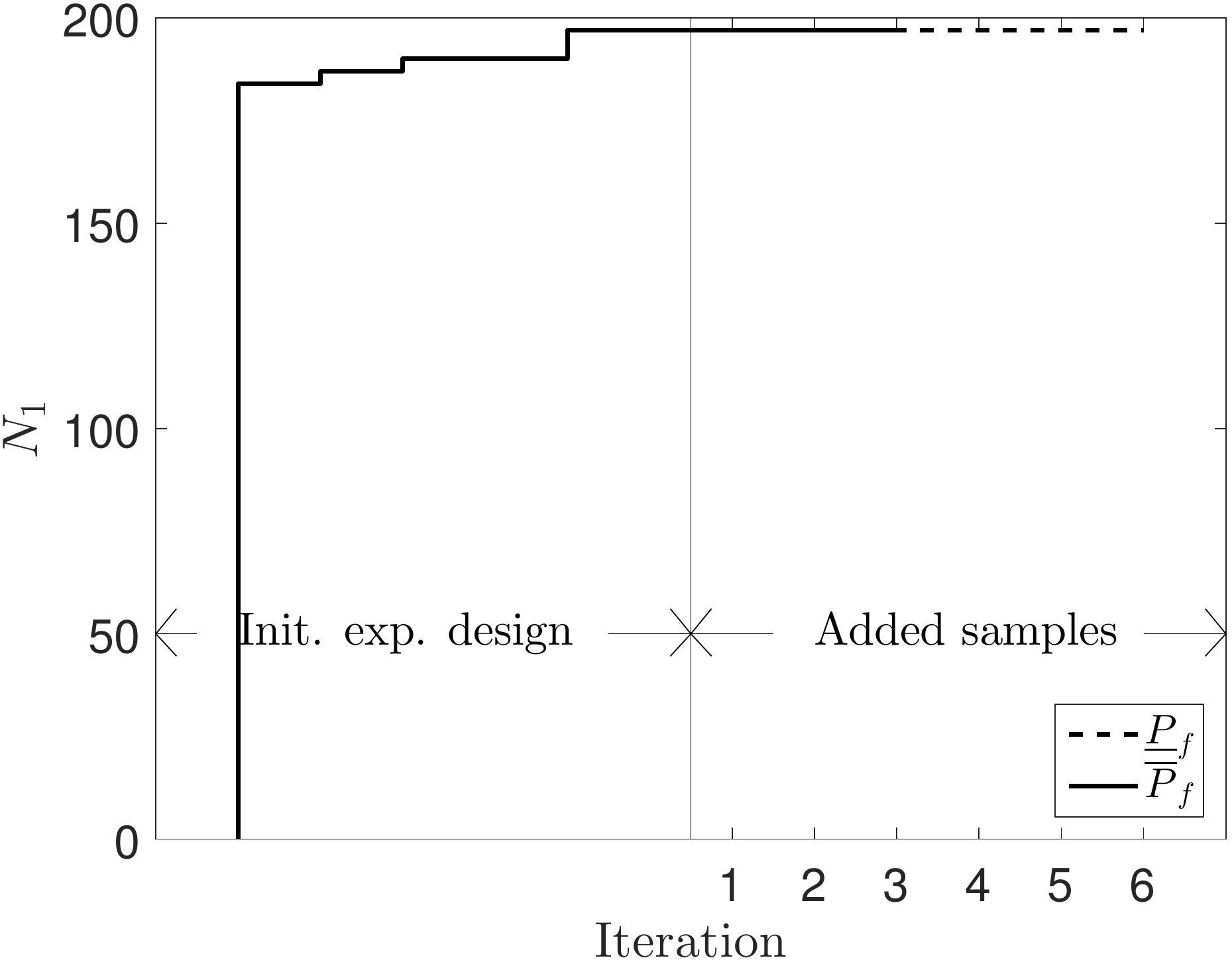}
}
\caption{Oscillator -- parametric p-boxes -- convergence of the ISRA \label{fig:osci:para}}
\end{figure}

\paragraph{Evolution of experimental designs}
As mentioned above, the evolution of the experimental design is different for the two types of p-boxes. For parametric p-boxes, the details are shown in Figure~\ref{fig:osci:para}. The size of $N_1$ is plotted as a function of the iteration in EGO in Figure~\ref{fig:osci:para:doe}. Interestingly, the major part of evaluations are required to estimate the first sample of EGO's initial experimental design. Only a few additional samples are evaluated in order to refine the conditional failure probability estimates. Hence, the recycling of limit-state function evaluations is an efficient tool to keep the computational costs at a low level.

In the same figure, the evolution of the boundary values of the failure probability are also given (see Fig.~\ref{fig:osci:para:pf}). Interestingly, the failure probabilities evolve with each iteration in EGO where at the same time $N_1$ remains constant. In other words, the limit-state surface $\cG=0$ is modelled accurately enough to estimate the extreme failure probabilities, \ie for different realizations $\vt\in\cD_{\vT}$. 

\paragraph{Effect of non-monotonicity on failure probability}
In this example, the two types of p-boxes have the same boundary curves, as seen in Table~\ref{tab:osci:variables}. In case the limit-state function was a monotone function, the failure probabilities would be identical. In other words, the same realizations of the input CDF would lead to one of the failure probability boundary values. 

In the case of the oscillator though, parametric and free p-box models result in significantly different failure probability bounds, as seen from Table~\ref{tab:osci:results}. In fact, the imprecise failure probabilities of the free p-box case encapsulate the ones of the parametric p-box case. As free p-boxes are more general than parametric p-boxes by definition, more extreme failure probabilities are possible. In this example, free p-boxes result in a $3.4$-times lower $\underline{P}_f$ and in a $1.8$-times larger $\overline{P}_f$ compared to the parametric p-boxes.

\subsection{Two-dimensional truss structure}

\subsubsection{Problem statement}
\begin{figure}[ht!]
\centering
\includegraphics[width=0.5\linewidth]{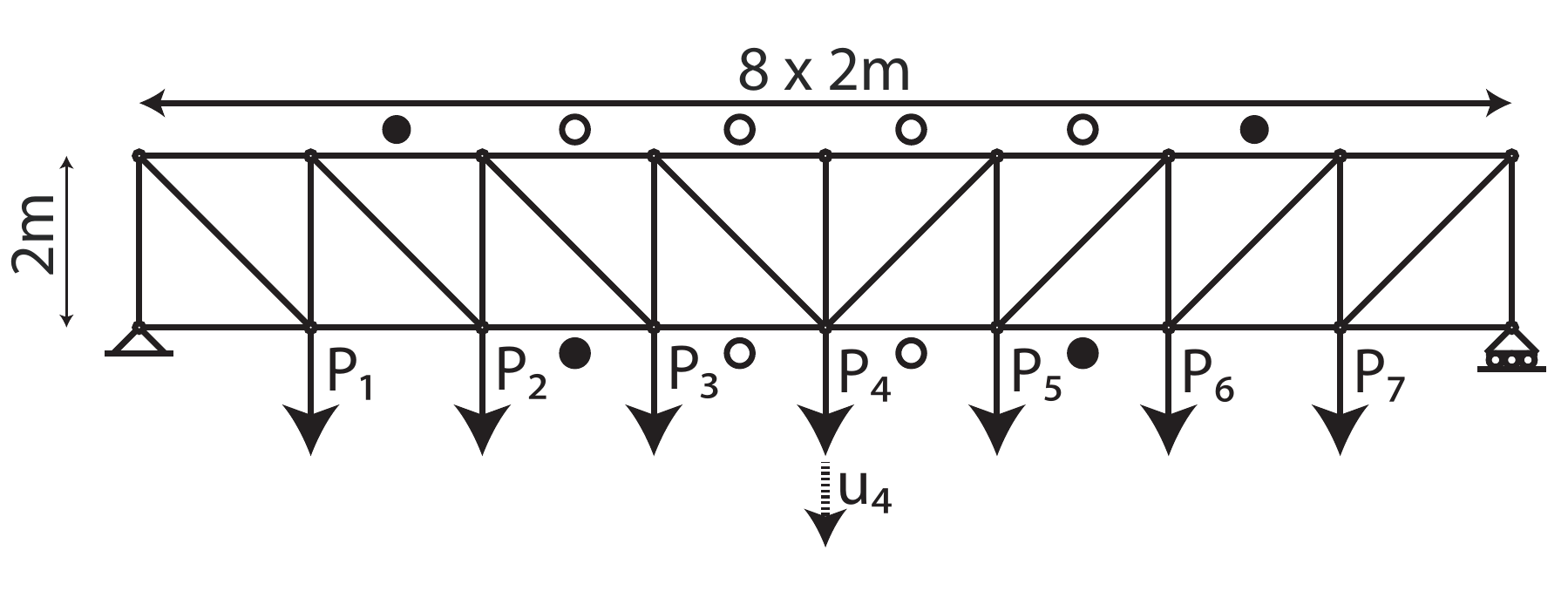}
\caption{\label{fig:truss:sketch} Frame structure -- sketch of the geometry and definition of the input variables}
\end{figure}

Hurtado \cite{Hurtado2013} introduced a two-dimensional linear-elastic truss structure, whose geometry is shown in Figure~\ref{fig:truss:sketch}. The truss is subjected to seven loads $P_i$ which are modelled with p-boxes. The parametric p-boxes are defined by lognormal distributions with mean value $\mu_{P_i}\in[95,105]~{\rm kN}$ and standard deviation $\sigma_{P_i}\in[13,17]~{\rm kN}$. The geometry of the structure and the material properties are assumed constant. The modulus of elasticity is $E=200\cdot10^9~{\rm Pa}$, whereas the cross section area varies along the different bars: $A=0.00535~{\rm m}^2$ for bars marked by $\bullet$, $A=0.0068~{\rm m}^2$ for bars marked by $\circ$, and $A=0.004~{\rm m}^2$ for the remaining bars.

In the following analysis, a second scenario is considered, where the former parametric p-boxes are modelled by free p-boxes. As in the other examples, the boundary curves of the p-boxes should coincide, hence:
\begin{eqnarray}
\underline{F}_{P_i}(p_i) &=& \min_{\mu_{P_i}\in[95,105]~{\rm kN}, \sigma_{P_i}\in[13,17]~{\rm kN}} F_{P_i}(p_i|\mu_{P_i},\sigma_{P_i}), \\
\overline{F}_{P_i}(p_i) &=& \max_{\mu_{P_i}\in[95,105]~{\rm kN}, \sigma_{P_i}\in[13,17]~{\rm kN}} F_{P_i}(p_i|\mu_{P_i},\sigma_{P_i}).
\end{eqnarray}

In the context of structural reliability analysis, the limit-state function is defined as
\begin{equation}
g_{truss}(\ve{p}) = 0.029~{\rm m} - u_4(\ve{p}),
\end{equation}
where $u_4$ is the deflection of the truss at midspan, as indicated in Figure~\ref{fig:truss:sketch}. Then, the failure probability describes the probability that the deflection of the truss exceeds $0.029~{\rm m}$, \ie $P_f=\Prob{u_4(\ve{P})\geq 0.029~{\rm m}}$.

\subsubsection{Analysis}

The deflection of the truss is computed by a \textsc{Matlab}-based finite element model (FEM), which is connected to the software framework \textsc{UQLab} \citep{MarelliICVRAM2014}. The FEM interprets each bar as a bar element, whereas the loads are modelled as point loads at the intersections of the corresponding bars as indicated in Figure~\ref{fig:truss:sketch}.

The ISRA settings are kept the same as in the previous examples. However, the initial experimental design is set to $N_1^0=N_2^0=12$ Latin-hypercube samples and $\epsilon_{EI}=10^{-5}$. For free p-boxes, the auxiliary distributions $\widetilde{X}_i$ are chosen as lognormal distributions with mean value $\mu_{\widetilde{P}_i}=100~{\rm kN}$ and standard deviation $\mu_{\widetilde{P}_i}=15~{\rm kN}$. 

\subsubsection{Results}

Table~\ref{tab:truss:results} summarizes the results in terms of failure probability estimates and number of model evaluations. The reference values are obtained by Monte Carlo simulation with $n=10^7$ samples, by evaluating the exact limit-state function, and by making use of the monotonicity property of the limit-state function. Using the proposed two-level algorithms, the failure probabilities are estimated accurately for both cases of parametric and free p-boxes. The numbers of FEM evaluations $N_1$ are low for both cases, too. Hence, the proposed multi-level meta-modelling techniques provide efficient algorithms for ISRA in this application example.

\begin{table}[ht!]
\centering
\caption{Truss -- results of the ISRA (reference solution obtained by Monte Carlo simulation with $n_{MC}=10^7$ samples)\label{tab:truss:results}}
\begin{tabular}{lccccc}
\hline
 & \multicolumn{2}{c}{Parametric p-box} && \multicolumn{2}{c}{Free p-box} \\
 & $\underline{P}_f$ & $\overline{P}_f$ && $\underline{P}_f$ & $\overline{P}_f$ \\
\hline
$P_{f,ref}$ & $2.49\cdot 10^{-4}$ & $7.62\cdot 10^{-2}$ && $2.21\cdot 10^{-4}$ & $9.22\cdot 10^{-2}$ \\
$\widehat{P}_f$ & $2.64\cdot 10^{-4}$ & $7.62\cdot 10^{-2}$ && $2.36\cdot 10^{-4}$ & $9.16\cdot 10^{-2}$\\
$N_1$ & $12+130=142$ & $12+129=141$&& $12+106=118$ & $12+106=118$\\
$N_2$ & $12+174=186$ & $12+40=52$ && $12+116=128$ & $12+192=204$\\
$N_2^*$ & $12+26=38$ & $12+4=16$ && - & - \\
\hline
\end{tabular}
\end{table}

In the case of parametric p-boxes, the minimum failure probability is obtained by $\mu_{P_i}=95~{\rm kN}$ and $\sigma_{P_i}=13~{\rm kN}$, whereas the maximum failure probability is obtained by $\mu_{P_i}=105~{\rm kN}$ and $\sigma_{P_i}=17~{\rm kN}$, $i=1,\ldots,7$, as one could expect. EGO is capable of identifying these two extreme cases and terminating the adaptive algorithm in this application in $174$ and $40$ iterations, which correspond to $N_2=186$ and $N_2=52$ samples in the EGO-related Kriging model, respectively. In this run, the number of iterations required to find those extremes is indicated in Table~\ref{tab:truss:results} by $N_2^*$. The values are lower than the final number of samples $N_2$, which are used to terminate the EGO algorithm. The additional samples beyond $N_2^*$ show the exploratory behaviour of the EGO algorithm in this application example and might be reduced by a more appropriate stopping criterion. 

Hurtado \cite{Hurtado2013} analysed the same truss structure for the parametric p-box case. However, due to its use of a Monte Carlo simulation on the second level of the algorithm, the published failure probability range is narrower, \ie $P_f\in\bra{3.2\cdot10^{-3},\, 2.37\cdot10^{-2}}$, than reported in Table~\ref{tab:truss:results}, which is arguably not conservative. This indicates the importance of using a proper optimization algorithm in order to find the extreme failure probabilities.

\section{Conclusions} \label{sec:conc}

In this paper, structural reliability analysis is discussed in the presence of deterministic, black-box performance functions. Traditionally, uncertainty in the input variables of such analyses is described by probability theory. In engineering practice however, a typical scenario consists of expensive data acquisition and thus limited data. Due to the sparsity of calibration data, a more general framework is required to characterize the input uncertainty.

In this paper, the input variables are modelled as imprecise probabilities accounting for aleatory (natural variability) and epistemic uncertainty (lack of knowledge). Probability-boxes (p-boxes) are one methodology of the class of imprecise probabilities that captures both types of uncertainty in a clear setting. In particular, two types of p-boxes are distinguished: free and parametric p-boxes. Free p-boxes model a variable by providing lower and upper bounds for the cumulative distribution function. Parametric p-boxes consist of a distribution function family whose parameters are modelled within intervals. 

In the context of p-boxes, structural reliability analyses are more complex than in the conventional probabilistic setting. Moreover, free and parametric p-boxes require different approaches. In order to cope with the complexity of imprecise structural reliability analyses, meta-models are used at different stages. In this paper, two two-level meta-modelling approaches are proposed for the cases of free and parametric p-boxes, respectively, making use of Kriging meta-models in combination with active learning algorithms. These approaches allow for an accurate estimation of the imprecise failure probabilities using only a limited number of runs of the performance function.

The capabilities of the two-level approaches are illustrated on a benchmark analytical function and two realistic engineering problems. In all examples, the proposed approaches are capable of estimating the imprecise failure probabilities in an efficient manner. In fact, the imprecise and conventional structural reliability analyses require a comparable number of runs of the performance function when using Kriging meta-models with adaptive experimental designs. This is of importance when the evaluations of the performance function are the dominating computational costs.


\bibliographystyle{chicago}
\bibliography{biblioISRA}

\begin{thebibliography}{}

\bibitem[\protect\citeauthoryear{Alvarez and Hurtado}{Alvarez and
  Hurtado}{2014}]{Alvarez2014}
Alvarez, D.~A. and J.~E. Hurtado (2014).
\newblock {An efficient method for the estimation of structural reliability
  intervals with random sets, dependence modeling and uncertain inputs}.
\newblock {\em Comput. Struct.\/}~{\em 142}, 54--63.

\bibitem[\protect\citeauthoryear{Au and Beck}{Au and Beck}{2001}]{Au2001}
Au, S. and J.~Beck (2001).
\newblock Estimation of small failure probabilities in high dimensions by
  subset simulation.
\newblock {\em Prob. Eng. Mech.\/}~{\em 16\/}(4), 263--277.

\bibitem[\protect\citeauthoryear{Au and Beck}{Au and Beck}{2003}]{Au2003}
Au, S. and J.~Beck (2003).
\newblock Subset simulation and its application to seismic risk based on
  dynamic analysis.
\newblock {\em J. Eng. Mech.\/}~{\em 129\/}(8), 901--917.

\bibitem[\protect\citeauthoryear{Bachoc}{Bachoc}{2013}]{Bachoc2012}
Bachoc, F. (2013).
\newblock {Cross Validation and Maximum Likelihood estimations of
  hyper-parameters of Gaussian processes with model misspecifications}.
\newblock {\em Comp. Stat. Data An.\/}~{\em 66}, 55--69.

\bibitem[\protect\citeauthoryear{Balesdent, Morio, and Brevault}{Balesdent
  et~al.}{2014}]{Balesdent2014}
Balesdent, M., J.~Morio, and L.~Brevault (2014).
\newblock Rare event probability estimation in the presence of epistemic
  uncertainty on input probability distribution parameters.
\newblock {\em Methodol. Comput. Appl. Probab.\/}, 1--20.

\bibitem[\protect\citeauthoryear{Bect, Ginsbourger, Li, Picheny, and
  Vazquez}{Bect et~al.}{2012}]{Bect2012}
Bect, J., D.~Ginsbourger, L.~Li, V.~Picheny, and E.~Vazquez (2012).
\newblock Sequential design of computer experiments for the estimation of a
  probability of failure.
\newblock {\em Stat. Comput.\/}~{\em 22\/}(3), 773--793.

\bibitem[\protect\citeauthoryear{Beer, Ferson, and Kreinovich}{Beer
  et~al.}{2016}]{BeerAPSSRA2016}
Beer, M., S.~Ferson, and V.~Kreinovich (2016).
\newblock Do we have compatible concepts of epistemic uncertainty?
\newblock In {\em Proc. {6th} {Asian-Pacific} {S}ymp. Struct. Reliab.
  {(APSSRA'2016)}, Shanghai}.

\bibitem[\protect\citeauthoryear{Ben-Haim}{Ben-Haim}{2006}]{BenHaim2006}
Ben-Haim, Y. (2006).
\newblock {\em {Info-gap decision theory: decisions under severe
  uncertainty}\/} (2nd editio ed.).
\newblock Academic Press, London.

\bibitem[\protect\citeauthoryear{Bichon, Eldred, Swiler, Mahadevan, and
  McFarland}{Bichon et~al.}{2008}]{Bichon2008}
Bichon, B.~J., M.~S. Eldred, L.~Swiler, S.~Mahadevan, and J.~McFarland (2008).
\newblock {Efficient global reliability analysis for nonlinear implicit
  performance functions}.
\newblock {\em AIAA Journal\/}~{\em 46\/}(10), 2459--2468.

\bibitem[\protect\citeauthoryear{Breitung}{Breitung}{1989}]{Breitung89}
Breitung, K. (1989).
\newblock Asymptotic approximations for probability integrals.
\newblock {\em Prob. Eng. Mech.\/}~{\em 4\/}(4), 187--190.

\bibitem[\protect\citeauthoryear{{De Angelis}, Patelli, and Beer}{{De Angelis}
  et~al.}{2015}]{DeAngelis2015}
{De Angelis}, M., E.~Patelli, and M.~Beer (2015).
\newblock {Advanced Line Sampling for efficient robust reliability analysis}.
\newblock {\em Structural Safety\/}~{\em 52}, 170--182.

\bibitem[\protect\citeauthoryear{{De Cooman}, Ruan, and Kerre}{{De Cooman}
  et~al.}{1995}]{DeCooman1995}
{De Cooman}, G., D.~Ruan, and E.~Kerre (1995).
\newblock {Foundations and applications of possibility theory}.
\newblock In {\em {Proc. FAPT~95}}, Singapore. World Scientific.

\bibitem[\protect\citeauthoryear{Dempster}{Dempster}{1967}]{Dempster1967}
Dempster, A.~P. (1967).
\newblock {Upper and lower probabilities induced by multivalued mapping}.
\newblock {\em Ann. Math. Stat.\/}~{\em 38\/}(2), 325--339.

\bibitem[\protect\citeauthoryear{Dubois and Prade}{Dubois and
  Prade}{1988}]{Dubois1988}
Dubois, D. and H.~Prade (1988).
\newblock {\em {Possibility theory: an approach to computerized processing of
  uncertainty}}.
\newblock New York: Plenum.

\bibitem[\protect\citeauthoryear{Dubourg, Sudret, and Deheeger}{Dubourg
  et~al.}{2013}]{Dubourg2013}
Dubourg, V., B.~Sudret, and F.~Deheeger (2013).
\newblock Metamodel-based importance sampling for structural reliability
  analysis.
\newblock {\em Prob. Eng. Mech.\/}~{\em 33}, 47--57.

\bibitem[\protect\citeauthoryear{Echard, Gayton, and Lemaire}{Echard
  et~al.}{2011}]{Echard2011}
Echard, B., N.~Gayton, and M.~Lemaire (2011).
\newblock {AK-MCS}: an active learning reliability method combining {K}riging
  and {M}onte {C}arlo simulation.
\newblock {\em Structural Safety\/}~{\em 33\/}(2), 145--154.

\bibitem[\protect\citeauthoryear{Echard, Gayton, Lemaire, and Relun}{Echard
  et~al.}{2013}]{Echard2013}
Echard, B., N.~Gayton, M.~Lemaire, and N.~Relun (2013).
\newblock {A combined Importance Sampling and Kriging reliability method for
  small failure probabilities with time-demanding numerical models}.
\newblock {\em Reliab. Eng. Syst. Safety\/}~{\em 111}, 232--240.

\bibitem[\protect\citeauthoryear{Eldred and Swiler}{Eldred and
  Swiler}{2009}]{Eldred2009}
Eldred, M.~S. and L.~P. Swiler (2009).
\newblock {Efficient algorithms for mixed aleatory-epistemic uncertainty
  quantification with application to radiation-hardened electronics part I :
  algorithms and benchmark results}.
\newblock Technical Report SAND2009-5805, Sandia National Laboratories.

\bibitem[\protect\citeauthoryear{Ferson and Ginzburg}{Ferson and
  Ginzburg}{1996}]{Ferson1996}
Ferson, S. and L.~R. Ginzburg (1996).
\newblock {Different methods are needed to propagate ignorance and
  variability}.
\newblock {\em Reliab. Eng. Syst. Safety\/}~{\em 54\/}(2-3), 133--144.

\bibitem[\protect\citeauthoryear{Ferson and Hajagos}{Ferson and
  Hajagos}{2004}]{Ferson2004}
Ferson, S. and J.~G. Hajagos (2004).
\newblock {Arithmetic with uncertain numbers: rigorous and (often) best
  possible answers}.
\newblock {\em Reliab. Eng. Syst. Safety\/}~{\em 85\/}(1-3), 135--152.

\bibitem[\protect\citeauthoryear{Gelman}{Gelman}{2006}]{Gelman2006}
Gelman, A. (2006).
\newblock {Prior distributions for variance parameters in hierarchical models
  (comment on article by Browne and Draper)}.
\newblock {\em Bayesian Anal.\/}~{\em 1\/}(3), 515--534.

\bibitem[\protect\citeauthoryear{Gelman, Carlin, Stern, and Rubin}{Gelman
  et~al.}{2009}]{Gelman2009}
Gelman, A., J.~Carlin, H.~Stern, and D.~Rubin (2009).
\newblock {\em Bayesian Data Analysis\/} (second ed.).
\newblock Chapman \& Hall/CRC.

\bibitem[\protect\citeauthoryear{Hohenbichler and Rackwitz}{Hohenbichler and
  Rackwitz}{1988}]{Hohenbichler1988}
Hohenbichler, M. and R.~Rackwitz (1988).
\newblock Improvement of second-order reliability estimates by importance
  sampling.
\newblock {\em J. Eng. Mech.\/}~{\em 114\/}(12), 2195--2199.

\bibitem[\protect\citeauthoryear{Hurtado}{Hurtado}{2013}]{Hurtado2013}
Hurtado, J.~E. (2013).
\newblock {Assessment of reliability intervals under input distributions with
  uncertain parameters}.
\newblock {\em Prob. Eng. Mech.\/}~{\em 32}, 80--92.

\bibitem[\protect\citeauthoryear{Jones, Schonlau, and Welch}{Jones
  et~al.}{1998}]{Jones1998}
Jones, D.~R., M.~Schonlau, and W.~J. Welch (1998).
\newblock {Efficient global optimization of expensive black-box functions}.
\newblock {\em J. Glob. Optim.\/}~{\em 13\/}(4), 455--492.

\bibitem[\protect\citeauthoryear{Kanno and Takewaki}{Kanno and
  Takewaki}{2006}]{Kanno2006}
Kanno, Y. and I.~Takewaki (2006, May).
\newblock {Robustness analysis of trusses with separable load and structural
  uncertainties}.
\newblock {\em Int. J. Solids Struct.\/}~{\em 43\/}(9), 2646--2669.

\bibitem[\protect\citeauthoryear{Koutsourelakis, Pradlwarter, and
  Schu\"eller}{Koutsourelakis et~al.}{2004}]{Koutsourelakis2004}
Koutsourelakis, P., H.~Pradlwarter, and G.~Schu\"eller (2004).
\newblock Reliability of structures in high dimensions, part {I}: algorithms
  and applications.
\newblock {\em Prob. Eng. Mech.\/}~{\em 19}, 409--417.

\bibitem[\protect\citeauthoryear{Lemaire}{Lemaire}{2009}]{Lemaire09}
Lemaire, M. (2009).
\newblock {\em Structural reliability}.
\newblock Wiley.

\bibitem[\protect\citeauthoryear{Marelli, {Lamas-Fernandes}, Sch\"obi, and
  Sudret}{Marelli et~al.}{2015}]{UQdoc_09_107}
Marelli, S., C.~{Lamas-Fernandes}, R.~Sch\"obi, and B.~Sudret (2015).
\newblock {UQLab} user manual -- {R}eliability analysis.
\newblock Technical report, Chair of Risk, Safety \& Uncertainty
  Quantification, ETH Zurich.
\newblock Report \# UQLab-V0.9-107.

\bibitem[\protect\citeauthoryear{Marelli and Sudret}{Marelli and
  Sudret}{2014}]{MarelliICVRAM2014}
Marelli, S. and B.~Sudret (2014).
\newblock {UQLab}: a framework for uncertainty quantification in {MATLAB}.
\newblock In {\em Proc. 2nd Int. Conf. on Vulnerability, Risk Analysis and
  Management {(ICVRAM2014)}, Liverpool, United Kingdom}.

\bibitem[\protect\citeauthoryear{McKay, Beckman, and Conover}{McKay
  et~al.}{1979}]{McKay1979}
McKay, M.~D., R.~J. Beckman, and W.~J. Conover (1979).
\newblock A comparison of three methods for selecting values of input variables
  in the analysis of output from a computer code.
\newblock {\em Technometrics\/}~{\em 2}, 239--245.

\bibitem[\protect\citeauthoryear{Mockus, Tiesis, and Zilinskas}{Mockus
  et~al.}{1978}]{Mockus1978}
Mockus, J., V.~Tiesis, and A.~Zilinskas (1978).
\newblock The application of {Bayesian} methods for seeking the extremum.
\newblock In {\em Towards Global Optimization}. Noth Holland, Amsterdam.

\bibitem[\protect\citeauthoryear{M\"oller and Beer}{M\"oller and
  Beer}{2004}]{Moller2004}
M\"oller, B. and M.~Beer (2004).
\newblock {\em {Fuzzy Randomness}}.
\newblock Springer.

\bibitem[\protect\citeauthoryear{Morio and Balesdent}{Morio and
  Balesdent}{2016}]{Morio2016}
Morio, J. and M.~Balesdent (2016, may).
\newblock {Estimation of a launch vehicle stage fallout zone with parametric
  and non-parametric importance sampling algorithms in presence of uncertain
  input distributions}.
\newblock {\em Aerosp. Sci. Technol.\/}~{\em 52}, 95--101.

\bibitem[\protect\citeauthoryear{Morio, Balesdent, Jacquemart, and
  Verg\'{e}}{Morio et~al.}{2014}]{Morio2014}
Morio, J., M.~Balesdent, D.~Jacquemart, and C.~Verg\'{e} (2014, December).
\newblock {A survey of rare event simulation methods for static input-output
  models}.
\newblock {\em Simul. Model. Pract. Theory\/}~{\em 49}, 287--304.

\bibitem[\protect\citeauthoryear{Oberguggenberger}{Oberguggenberger}{2014}]{Oberguggenberger2014}
Oberguggenberger, M. (2014).
\newblock {Analysis and computation with hybrid random set stochastic models}.
\newblock {\em Structural Safety\/}~{\em 52\/}(Part B), 233--243.

\bibitem[\protect\citeauthoryear{Santner, Williams, and Notz}{Santner
  et~al.}{2003}]{Santner2003}
Santner, T.~J., B.~J. Williams, and W.~I. Notz (2003).
\newblock {\em {The Design and Analysis of Computer Experiments}}.
\newblock Springer, New York.

\bibitem[\protect\citeauthoryear{Sch\"obi and Sudret}{Sch\"obi and
  Sudret}{2015a}]{SchoebiESREL2015}
Sch\"obi, R. and B.~Sudret (2015a).
\newblock Imprecise structural reliability analysis using {PC-Kriging}.
\newblock In {\em Proc. 25th European Safety and Reliability Conference
  (ESREL2015), Zurich, Switzerland}.

\bibitem[\protect\citeauthoryear{Sch\"obi and Sudret}{Sch\"obi and
  Sudret}{2015b}]{SchoebiICASP2015}
Sch\"obi, R. and B.~Sudret (2015b).
\newblock Propagation of uncertainties modelled by parametric p-boxes using
  sparse polynomial chaos expansions.
\newblock In {\em Proc. 12th~Int. Conf. on Applications of Stat. and Prob. in
  Civil Engineering (ICASP12), Vancouver, Canada}.

\bibitem[\protect\citeauthoryear{Sch\"obi and Sudret}{Sch\"obi and
  Sudret}{2016}]{SchoebiJCP2016}
Sch\"obi, R. and B.~Sudret (2016).
\newblock Uncertainty propagation of p-boxes using sparse polynomial chaos
  expansions.
\newblock {\em J. Comput. Phys.\/}, submitted.

\bibitem[\protect\citeauthoryear{Sch\"obi, Sudret, and Marelli}{Sch\"obi
  et~al.}{2016}]{SchobiASCE2015}
Sch\"obi, R., B.~Sudret, and S.~Marelli (2016).
\newblock Rare event estimation using {Polynomial-Chaos-Kriging}.
\newblock {\em ASCE-ASME J. Risk Uncertain. Eng. Syst. Part A Civ. Eng.\/},
  D4016002.

\bibitem[\protect\citeauthoryear{Schueremans and {Van Gemert}}{Schueremans and
  {Van Gemert}}{2005}]{Schueremans2005}
Schueremans, L. and D.~{Van Gemert} (2005).
\newblock {Benefit of splines and neural networks in simulation based
  structural reliability analysis}.
\newblock {\em Structural Safety\/}~{\em 27\/}(3), 246--261.

\bibitem[\protect\citeauthoryear{Shafer}{Shafer}{1976}]{Shafer1976}
Shafer, G. (1976).
\newblock {\em {A mathematical theory of evidence}}.
\newblock Princeton, NJ: Princeton University Press.

\bibitem[\protect\citeauthoryear{Zadeh}{Zadeh}{1978}]{Zadeh1978}
Zadeh, L. (1978).
\newblock {Fuzzy sets as a basis for a theory of possibility}.
\newblock {\em Fuzzy Sets Syst.\/}~{\em 1}, 3--28.

\bibitem[\protect\citeauthoryear{Zhang}{Zhang}{2012}]{Zhang2012a}
Zhang, H. (2012, September).
\newblock {Interval importance sampling method for finite element-based
  structural reliability assessment under parameter uncertainties}.
\newblock {\em Structural Safety\/}~{\em 38}, 1--10.

\bibitem[\protect\citeauthoryear{Zhang, Mullen, and Muhanna}{Zhang
  et~al.}{2010}]{Zhang2010}
Zhang, H., R.~L. Mullen, and R.~L. Muhanna (2010).
\newblock Interval {Monte Carlo} methods for structural reliability.
\newblock {\em Structural Safety\/}~{\em 32}, 183--190.

\bibitem[\protect\citeauthoryear{Zhang, Jiang, Han, Hu, and Yu}{Zhang
  et~al.}{2014}]{Zhang2014}
Zhang, Z., C.~Jiang, X.~Han, D.~Hu, and S.~Yu (2014).
\newblock {A response surface approach for structural reliability analysis
  using evidence theory}.
\newblock {\em Adv. Eng. Softw.\/}~{\em 69}, 37--45.

\bibitem[\protect\citeauthoryear{Zhang, Jiang, Wang, and Han}{Zhang
  et~al.}{2015}]{Zhang2015}
Zhang, Z., C.~Jiang, G.~Wang, and X.~Han (2015).
\newblock {First and second order approximate reliability analysis methods
  using evidence theory}.
\newblock {\em Reliab. Eng. Syst. Safety\/}~{\em 137}, 40--49.

\end{thebibliography}

\end{document}